\documentclass[final,5p,times,twocolumn,a4paper]{elsarticle}

\usepackage{hyperref}
\usepackage{url}
\usepackage{graphicx}
\usepackage[update,prepend]{epstopdf}
\usepackage{amsmath}
\usepackage{siunitx}
\usepackage{color}
\usepackage{booktabs}

\usepackage{amssymb}
\usepackage{lineno}
\usepackage{url}
\usepackage{siunitx}
\usepackage{xspace}
\usepackage{gensymb}

\usepackage[colorinlistoftodos]{todonotes}

\journal{Nucl. Instrum. Methods Phys. Res., Sect. A}

\begin{document}
\begin{frontmatter}

\title{Compact, directional neutron detectors capable of high-resolution nuclear recoil imaging}


\author[add1,add2]{I.~Jaegle}
\author[add1]{P.~M.~Lewis}
\author[add4]{M.~Garcia-Sciveres}
\author[add1]{M.~T.~Hedges}
\author[add3]{T.~Hemperek}
\author[add3]{J.~Janssen} 
\author[add4]{Q.~Ji} 
\author[add3]{D.-L. Pohl} 
\author[add1]{S.~Ross}
\author[add1]{J.~Schueler}
\author[add1]{I.~Seong}
\author[add1]{T.~N.~Thorpe}
\author[add1]{S.~E.~Vahsen\corref{cor1}}
\cortext[cor1]{Corresponding author. Tel.: +1 808 956 2985.}
\ead{sevahsen@hawaii.edu}

\address[add1]{University of Hawaii, Department of Physics and Astronomy, 2505 Correa Road, Honolulu, HI 96822, USA}\address[add2]{University of Florida, Department of Physics, P.O. Box 118440, Gainesville, FL 32611, USA}
\address[add3]{University of Bonn, Institute of Physics, Nu{\ss}allee 12, 53115 Bonn, Germany}
\address[add4]{Lawrence Berkeley National Laboratory, 1 Cyclotron Road, Berkeley, CA 94720, USA}


\begin{abstract}
We report on the design, production, and performance of compact 40-cm$^3$ Time Projection Chambers (TPCs) that detect fast neutrons by measuring the three-dimensional (3D) ionization distribution of nuclear recoils in $^4$He:CO$_2$ gas at atmospheric pressure. We use these detectors to characterize the fast-neutron flux inside the Belle II detector at the SuperKEKB electron--positron collider in Tsukuba, Japan, where the primary design constraint is a small form factor. 
We find that the TPCs meet or exceed all design specifications, and are capable of measuring the 3D surface shape and charge density profile of ionization clouds from nuclear recoils and charged tracks in exquisite detail. Scaled-up detectors based on the detection principle demonstrated here may be suitable for directional dark matter searches, measurements of coherent neutrino-nucleus scattering, and other experiments requiring precise detection of neutrons or nuclear recoils.
\end{abstract}


\end{frontmatter}


\section{Introduction}\label{introduction}

Nuclear recoils provide a unique means to detect the elastic scattering of fast neutrons~\cite{Vahsen:2014fba}, the coherent elastic scattering of neutrinos~\cite{Akimov:2017ade}, and possibly, if weakly interacting massive particles (WIMPs) are the constituents of dark matter, WIMP-nucleon scattering~\cite{Goodman:1984dc}. Directional detection of nuclear recoils is therefore desirable in a broad range of scientific and technological fields, including neutron imaging, homeland security, nuclear physics, particle physics, and space sciences. For instance, directional recoil detection may be a key ingredient for unambiguously demonstrating the cosmological origin of a tentative dark matter signal~\cite{Ahlen:2009ev, Mayet:2016zxu, Battat:2016pap}, and for penetrating the so-called neutrino floor~\cite{OHare:2016pjy} that is likely to impede conventional (non-directional) dark matter searches in the future. Micro pattern gaseous detectors~\cite{Sauli:1997qp,Giomataris:1995fq} have made it feasible to read out the 3D ionization density in time projection chambers (TPCs) with high spatial resolution. This has enabled the reconstruction of the 3D direction and energy of short, mm-scale nuclear recoils~\cite{Vahsen:2014fba}. 

Here, we report on the BEAST TPCs, a new generation of directional nuclear recoil detectors, consisting of miniature gas time projection chambers~\cite{Nygren:1978rx} where the ionization is avalanche-multiplied with Gas Electron Multipliers (GEMs)~\cite{Sauli:1997qp} and detected with the ATLAS FE-I4B Pixel Application Specific Integrated Circuit (ASIC)~\cite{GARCIASCIVERES2011S155}. Although the principle of detection we use here is widely applicable, we have adopted a highly compact form factor, optimized for measuring the directional distribution and energy spectrum of fast neutron recoils inside the Belle II detector~\cite{Abe:2010gxa} at the SuperKEKB collider~\cite{Ohnishi:2013fma} in Japan. 

\begin{figure}[h]
\centering
\includegraphics[width=.8\columnwidth]{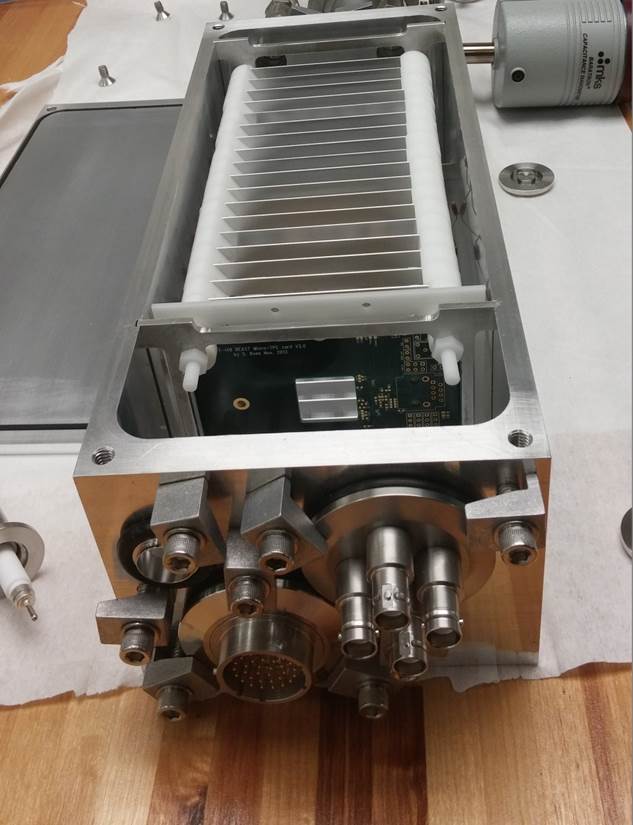}
\caption{A prototype BEAST TPC fast neutron detector, with $60~{\rm cm}^3$ sensitive volume. We built two such detectors, and deployed one during SuperKEKB commissioning Phase 1. Eight final detectors with minor geometrical modifications and improvements were also constructed.}
\label{fig:neutron_tpc}
\end{figure}

Measurements of neutrons at Belle II are motivated by high expected neutron rates at SuperKEKB, which may eventually shorten the lifetime and degrade the performance of Belle II detector systems. It is therefore important to validate the neutron background simulations so that shielding can be added where required before the collider reaches full luminosity. We have constructed two BEAST TPC prototypes (Fig.~\ref{fig:neutron_tpc}) and eight final detectors (Fig.~\ref{fig:final_insides}) to perform these measurements. 

In 2016, during the first phase of SuperKEKB operations (Phase 1), we deployed one prototype and one final detector near the interaction point before the Belle II detector was in place. Measurements from this run have already been reported~\cite{Lewis:2018ayu,phdhedges}. For the second phase of SuperKEKB operations (Phase 2), from February to July 2018, we deployed eight final TPCs in temporary unused volume inside the Belle II detector, which was in its final location at the interaction point. This allowed us to perform {\it in situ} measurement of neutrons incident on the Belle II detector systems, generated by showers from off-orbit accelerator beam particles hitting the accelerator beampipe. Measurements from this run will be reported separately. We have recently installed six of the final TPCs in the SuperKEKB tunnel for the final data-taking phase of SuperKEKB, which began in March 2019.

We report here on the design of the BEAST TPCs themselves, on improvements in the readout firmware that were required to achieve the reconstruction of nuclear recoils and enabled a veto of triggers generated by electron recoils, and on the performance of the detectors measured with particle sources during production. The angular and energy resolution for nuclear recoils, as a function of recoil energy, will be reported separately. Selected preliminary results can be found in Refs.~\cite{phdhedges,phdthorpe}. 

\section{Basic detector design}\label{basic_design}

The BEAST TPCs are based on a previous generation of smaller detectors developed by our group, which are described in Ref.~\cite{Vahsen:2014fba}. We refer the reader to that reference for a more general and detailed discussion of TPC charge readout with GEMs and a pixel chip. The functionality of the pixel chip is summarized in Section~\ref{sec:daq}. In this section we describe only the details sufficient to support the following discussion. 

\subsection{Physical description}
Our basic design consists of a volume of gas inside a uniform electric field maintained by a field cage. Ionized charge in the gas drifts to the anode of the field cage and passes through a pair of GEMs, each of which avalanche-multiplies the charge by a factor of order 100. We record the resulting avalanche charge with high resolution in both time and space with a pixel chip. 

For calibrations and performance monitoring, we embed between one and three $^{210}$Po alpha sources on the field cage so that the alphas traverse the entire width of the sensitive volume.

\subsection{Improvements over early prototypes}

The BEAST TPCs represent a factor of 60 (prototype) to 40 (final detector) scale-up in target volume over detectors of similar design previously constructed by our group \cite{Vahsen:2014fba}. The larger target volume is achieved by increasing the drift length to 10 (15) cm, and using a larger-area ATLAS pixel chip~\cite{GARCIASCIVERES2011S155}. The longer drift length necessitates a field cage to maintain drift field uniformity, and requires custom DAQ firmware to read out large events that span more than 16 consecutive pixel chip clock cycles --- far from the regular use case of the ATLAS pixel chip, but made possible by a special chip operation mode implemented by the chip designers with an eye to our application. Only a small fraction of the pixel chip surface consists of conductive bump pads. In order to increase the conductive area that can collect drift charge, we deposit a pixelized metal layer onto the pixel chip. In previous detectors, we deposited the layer ourselves, which led to non-uniform response~\cite{Vahsen:2014fba}. We now use a commercial vendor and see no such non-uniformity. 

\subsection{3D reconstruction}
\begin{figure*}
\centering
\includegraphics[height=6cm]{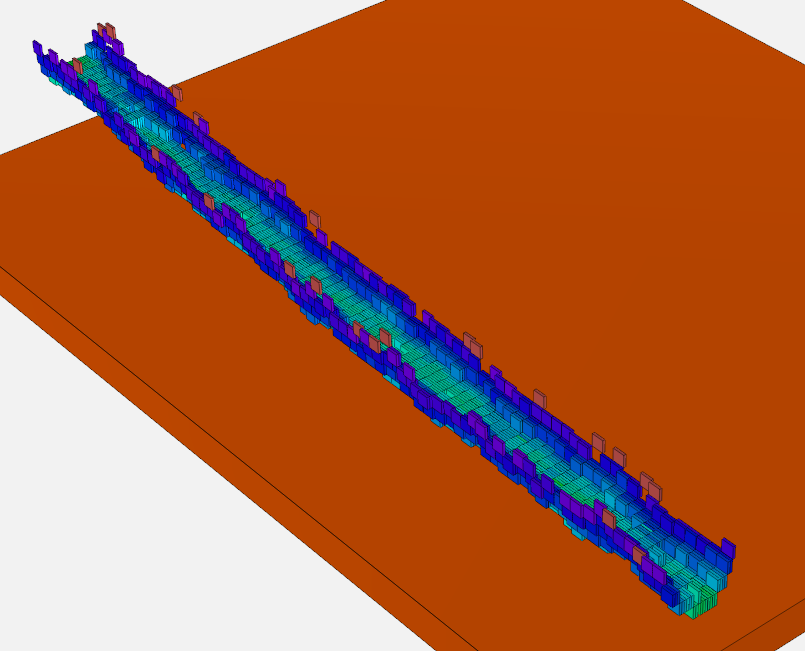}
\includegraphics[height=6cm]{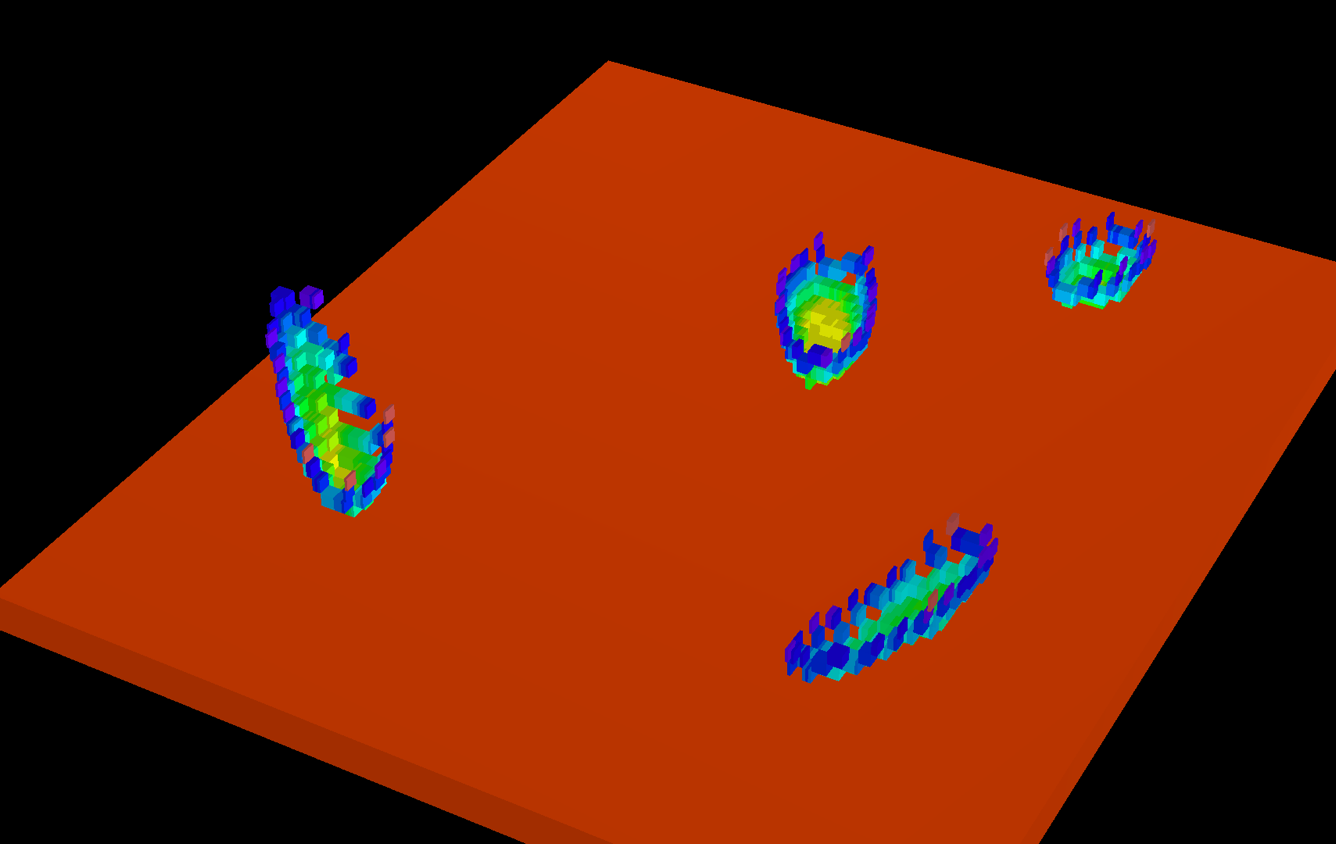}
\caption{(color online) Example of reconstructed 3D ionization distributions. Left: an alpha particle track from one of the calibration sources. Right: four superimposed fast-neutron recoil tracks. In both figures, the color of each box indicates the amount of ionization recorded in each 250 $\times$ \SI{50}{\micro\meter} pixel with a signal above threshold. Green is higher ionization density, blue is lower. The vertical position of each box is assigned using the relative timing of threshold crossing for each pixel. Because the pixel integration time is longer than the typical charge cloud size, only one entry per horizontal position is measured, leading to the canoe shape of the reconstructed tracks. The large orange shape shows the dimensions of the pixel chip for scale.}
\label{fig:event_displays}
\end{figure*}

The digitized data from the pixel chip consists of row, column, relative time of threshold crossing, and TOT (time-over-threshold, a measurement of integrated charge) for each pixel that crossed threshold in the event, visualized in Fig.~\ref{fig:event_displays}. We assign physical coordinates to the pixel hits so that $x$ corresponds to the column, $y$ to the row, and $z$ to the relative threshold-crossing time of the hit. We use a calibration to convert TOT to charge, and therefore our data now consists of charge measured in discrete positions in 3D space. 

To obtain the track direction, we fit the charge distribution in space with a line to obtain the polar angle $\theta$ with respect to the drift axis, and the azimuthal angle $\phi$ with respect to the $x$ axis of the chip. 

\section{Detector design optimization}

\label{sec:opti}

The specific design of the BEAST TPCs is dictated by several application-specific constraints. We categorize these as Environmental, Directionality, Miniaturization or Uniformity constraints. In this section we describe these constraints and their effect on our design choices for the BEAST TPCs. 

\subsection{Environmental constraints}
The Environmental constraints can be summarized as requiring that the BEAST TPCs operate stably \textit{in situ} (inside Belle II) and that they do not interfere with the stable operation of neighboring Belle II subdetectors. The primary concerns are gas safety, operation in the 1.5~T Belle II solenoid field, electromagnetic interference, and the use of high voltage (HV).

\subsubsection{Drift gas\label{sec:gas}}
We seek a suitable drift gas that is inert and inexpensive. To this end, we evaluate gas mixtures containing $^4$He nuclei as neutron targets and an additional inert quenching gas. To find the optimum gas, we develop the following figure of merit (FOM$_{\textrm{gas}}$): 
\begin{equation}
{\rm FOM}_{\textrm{gas}} = \frac{E_{\textrm{ionization}}G \varepsilon}{D_t^2 W},
\end{equation}
where $E_{\textrm{ionization}}$ is the ionization energy of the recoil, $G$ is the intrinsic gas gain, $\varepsilon$ is the neutron-nucleus interaction probability, $D_t$ is the transverse diffusion, and $W$ is the work function of the gas. We calculate $E_{\textrm{ionization}}$ using SRIM~\cite{SRIM} for the maximum momentum transfer by a 1~MeV neutron to a $^1$H or $^4$He nucleus due to elastic scattering. Because the pixel chip digitizes hits with a period of 25~ns, we require a drift velocity of \SI{10}{\micro\meter}/ns in order to achieve the same spatial quantization (\SI{250}{\micro\meter} in the drift direction) as the largest pixel dimension. This choice is motivated further in section \ref{field_optimization}. We then use Magboltz~\cite{magboltz,Biagi:1999nwa} to calculate the transverse diffusion $D_t$ at the drift field value that gives this drift velocity. We calculate the intrinsic gain $G$ using Magboltz for a GEM field of 0.1~MV/cm. We determine the interaction probability $\varepsilon$ using GEANT4~\cite{Agostinelli:2002hh}. 

\begin{figure}[h!]
  \centering
  \includegraphics[width=\columnwidth]{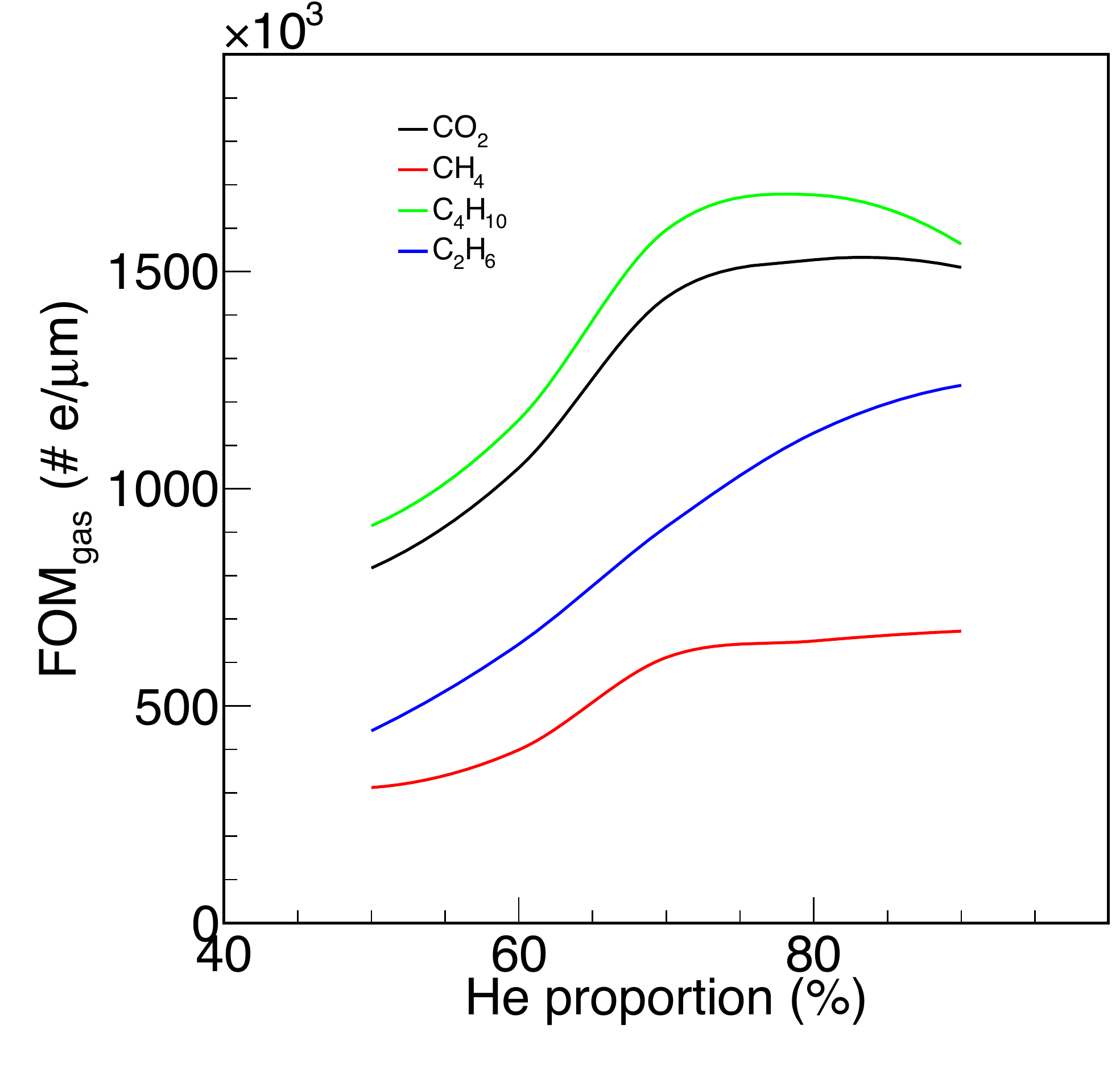}
  \caption{Gas figure of merit (FOM$_{\textrm{gas}}$) values versus $^4$He proportion for 0.6~MeV $^4$He recoils in different gas mixtures: $^4$He:CO$_2$ (black), He:CH$_4$ (red), $^4$He:C$_4$H$_{10}$ (green), $^4$He:C$_2$H$_6$ (blue) at 1~atm.}
  \label{fig:fom}
\end{figure}

Figure~\ref{fig:fom} shows FOM$_{\textrm{gas}}$ values for different gas mixtures versus the $^4$He proportion at 1~atm. While isobutane (C$_4$H$_{10}$) has the highest FOM$_{\textrm{gas}}$ value, it is flammable. We instead choose the runner-up He:CO$_2$, which is inert. The FOM$_{\textrm{gas}}$ values plateau around a Helium proportion of 70\%, which we choose to maximize the amount of the quenching gas CO$_2$. For this 70:30 He:CO$_2$ mixture, the required drift velocity of \SI{10}{\micro\meter}/ns occurs at a drift field of 530~V/cm. We select this gas and drift field for the BEAST TPCs.

\subsubsection{Vessel material}
We look for a vessel material that has very low magnetic permeability (to minimize distortions in the 1.5~T axial Belle II solenoid field), low activation, low outgassing (to minimize gas contamination inside the vessel), and low hydrogen content (to minimize proton backgrounds). We therefore select aluminum 6063 as our vessel material, as it is inexpensive, readily available, and satisfies all of these requirements. 

\subsection{Directionality constraints}
The baseline requirement of TPCs in BEAST is that they must be able to uniquely identify neutron recoils and measure their direction in 3D space over a broad range of expected recoil energies. This requirement influences a large number of design choices.

\subsubsection{Gain}
We use GEMs to amplify the ionization charge of neutron recoils. We use a pair of GEMs, each of which has a gain of order 100. Due to the limited dynamic range of the charge scale of the chip and other factors there is an optimum gain at which the charge recovery efficiency is maximized. We define a gain figure of merit:
\begin{equation}
{\rm FOM}_{\textrm{gain}} =  \frac{Q_{\textrm{recovered}}}{Q_{\textrm{all}}},
\label{eqn:fom_gain}
\end{equation}
where $Q_{\textrm{all}}$ is the total charge of a track after drift and amplification and just before arriving at the chip, and $Q_{\textrm{recovered}}$ is the charge collected by the chip, which is $Q_{\textrm{all}}$ minus the charge lost below the TOT threshold and due to TOT saturation. We expect FOM$_{\textrm{gain}}$ to be suppressed at low gains due to charge lost below threshold and also at high gains due to charge loss through TOT scale saturation. 

\begin{figure} 
  \begin{center}
    \includegraphics[width=\columnwidth]{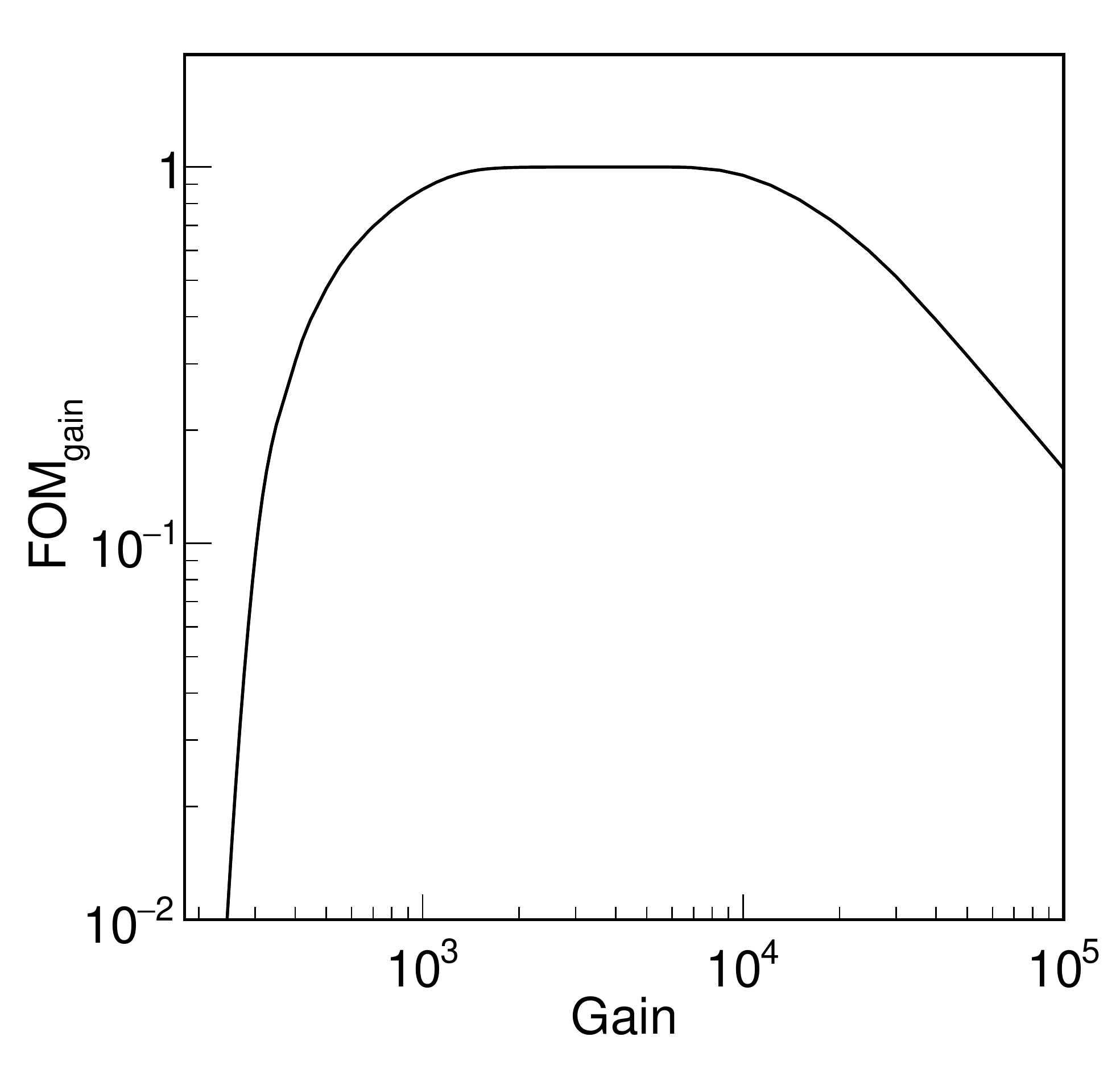}
    \caption{Gain figure of merit (FOM$_{\textrm{gain}}$) value for the ionization charge from a 300~keV $^4$He recoil after drifting 8~cm.}
    \label{fig:fom_gain}
  \end{center}
\end{figure}

To calculate the optimum gain, we simulate the ionization, drift, amplification and digitization of a 300~keV $^4$He recoil deposited 8~cm above the top GEM for a range of gains. The results, shown in Fig.~\ref{fig:fom_gain}, indicate that the optimum gain is in the range 1500 to 7000 and we choose GEM voltages to achieve a gain of 1500.

\subsubsection{\label{field_optimization}Pixel chip and drift velocity}
We require that the recoil directionality is limited by diffusion rather than the spatial and temporal resolution of the readout. Because the diffusion is determined by the drift velocity, the choices of readout technology and drift velocity are coupled. Therefore we first select a high-resolution pixel chip, the FE-I4B~\cite{Backhaus:2013}. The FE-I4B has a 2~cm $\times$ 1.68~cm active area consisting of 80 $\times$ 336 pixels measuring 250 $\times$ \SI{50}{\micro\meter} and digitizing in time bins of 25~ns. The chip capabilities and its use in the BEAST TPCs are discussed further in Section~\ref{sec:daq}.

With a high-resolution pixel chip, we now can optimize the drift velocity. If the drift velocity is too low, diffusion obscures the directionality of the drifted charge. Additionally, longer drift times lead to more charge loss through re-absorption with impurities in the gas, such as oxygen. Conversely, if the drift velocity is too high, the angle of the track with respect to the plane of the pixel chip cannot be measured. We define the longitudinal resolution as the time resolution multiplied by the drift velocity. To select our optimal drift velocity, we choose the maximum velocity that yields a longitudinal resolution no worse than the lateral resolution, in other words: 25\,ns$\cdot v_d$ = \SI{250}{\micro\meter}, so $v_d$ = \SI{10}{\micro\meter}/ns. Also see the discussion related to drift velocity in Section~\ref{sec:gas}.

\subsection{Miniaturization constraints}
We choose a vessel form factor that maximizes internal volume within the allotted Belle II space, which is roughly $14\times 16\times 40$\,cm$^3$. Miniaturization to this scale creates potential problems with gas quality and HV protection that we discuss further in this section. 

\subsubsection{Internal materials}
The internal components of the TPCs must satisfy the same specifications as the vessel material, including low magnetic permeability, low activation, low outgassing, and low hydrogen content. Due to the small gas volume of the TPCs, the ratio of surface area of all internal components to the volume of gas is unusually high. Consequently, minimizing outgassing is the most pressing of these constraints. 

For the field cage rings we select the same material as the vessel, aluminum 6063. We use Acetal, an electrically insulating thermoplastic, to hold the field cage together and space the rings. For HV protection we use Kapton, a low-outgassing insulator with a dielectric strength of 77~kV/mm.

The pixel chip, its associated printed circuit board, components, and wiring are not optimized for their outgassing properties.

\subsubsection{High voltage insulation}
The breakdown field strength of a 70:30 mixture of He:CO$_2$ at atmospheric pressure is 8~kV/cm. Given that a field cage voltage of 8~kV is needed to achieve the target drift velocity, and that there is a 1~cm gap between the field cage and vessel, we expect HV discharges between the field cage and the vessel to be a significant risk. 

\begin{figure*}
\centering
\includegraphics[width=\columnwidth]{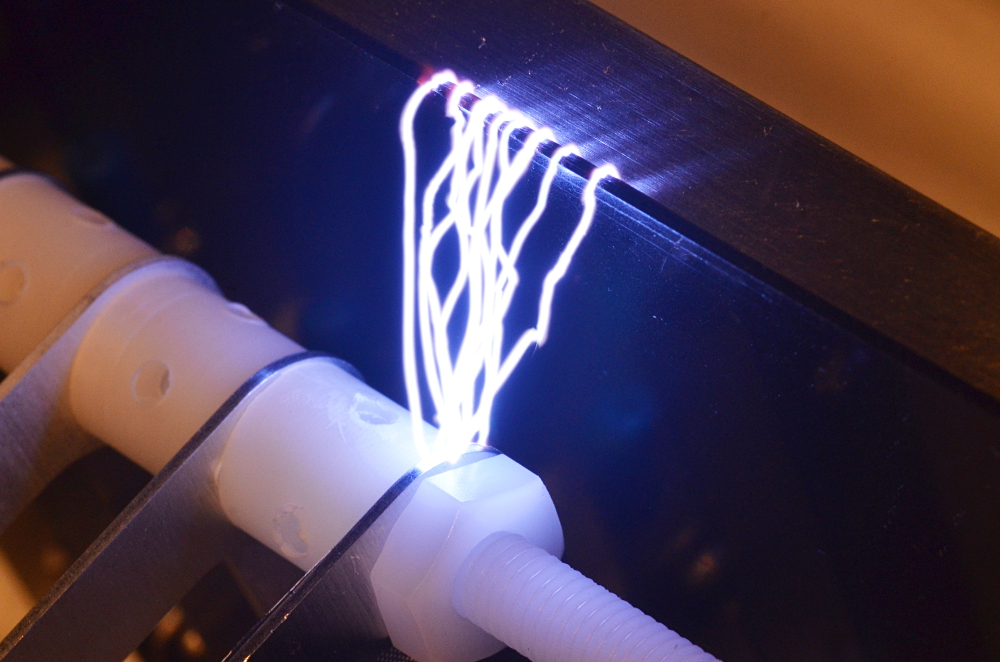} 
\includegraphics[width=\columnwidth]{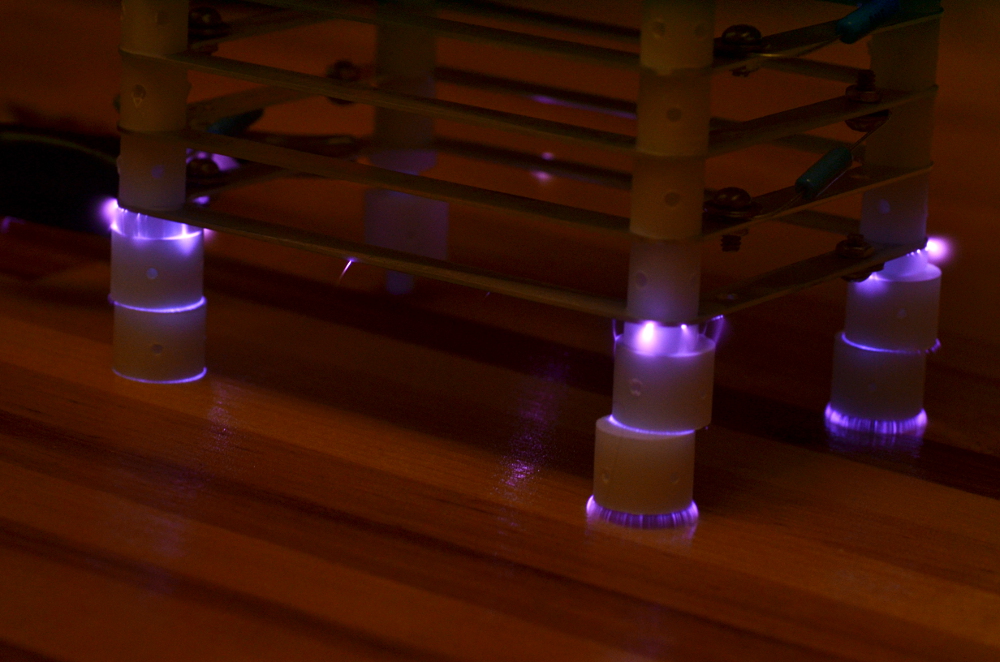}
\caption{Left: a long time exposure showing multiple discharges between the top field cage ring (the ``cathode mesh'', foreground) and the vessel (background) at 30~kV. The discharges find the shortest path around, rather than through, the Kapton sheet (dark plane). Right: a long time exposure showing corona discharges from the cathode mesh (bottom). The field cage is shown supported by antistatic Acetal spacers, with the cathode directly above a table. The field cage rings in this test were machined using CNC milling, the method that left the fewest defects and thus the least amount of corona activity.}
\label{fig:discharge}
\end{figure*}

In order to investigate HV discharges we perform a series of open-air tests up to 30~kV (roughly equivalent to 8~kV in He:CO$_2$). We find that inserting 0.5~mm-thick Kapton sheets between the field cage and the vessel increases the voltage we can apply to the field cage before discharges occur. However, the discharges still exploit the shortest path between field cage and vessel at 30~kV (Fig.~\ref{fig:discharge}, left). Consequently, we add a second layer of insulation: a 1~mil deposition of Parylene C~\cite{parylenec} on the inside surface of the vessel, which increases the breakdown voltage by 5.3~kV. As an additional benefit, Parylene C has low permeability to oxygen and other contaminating gases that may desorb from the vessel walls, effectively sealing the gas volume from a large source of outgassing. While a thicker layer of Parylene C would have provided a greater safety margin for electrical discharges, the price was prohibitive.
  
While the field cage is outside the vessel we observe no discharges. However, we do see persistent corona generated by concentrated electric fields around minor machining defects on the field cage rings, particularly the ``cathode mesh'', the final ring, which is furthest from ground (Fig.~\ref{fig:discharge}, right). While corona itself is not a concern, we expect an increased likelihood of discharges from these defects and therefore use the coronas to investigate machining quality on the field cage rings.  We test three different manufacturing processes for the field cage rings: laser cut, water jet and Computer Numerical Control (CNC) milling with a \SI{25}{\micro\meter} precision cut. We find that CNC gives the most reliable results, minimizing both the number and the power of the corona discharges. 

For the BEAST TPCs we have adopted all of these solutions: 0.5~mm Kapton sheets surrounding the field cage, 1~mil Parylene C coating on the interior vessel walls, and CNC-milled field cage rings.

\subsection{Uniformity constraints}
The final design constraint dictates that the TPC performance is spatially uniform and stable with time. Specifically, the TPCs must be able to detect and measure directionality for a broad range of neutron recoil energies with consistent gain and uniformity of response in the readout ($x-y$) plane. To achieve a response that is uniform also in the drift direction, $z$, distortions introduced by the drift field should be small compared to the pixel chip resolution. Finally, detector performance should be stable and sufficient to meet our physics goals throughout six months of operation in BEAST. 

\subsubsection{Field uniformity}
We design the field cage to maintain a uniform electric field parallel to the $z$-axis. Drifting electrons in this field follow electric field lines. Any component of the field that is transverse to the $z$ axis will therefore distort drifted tracks. A suitable field cage design will therefore minimize the transverse field $E_r(x,y,z)$ and maintain a constant $z$ component of the field $E_z(x,y,z)$ throughout the volume. 

We can approximate the maximum transverse displacement of a drifting electron in the field cage as
\begin{equation}
\Delta r(x,y) = \int_{z=0}^{z_{max}/2} \frac{E_r(x,y,z)}{E_z(x,y,z)} dz,
\label{eqn:transverse_displacement}
\end{equation}
where $z$ is the height of the electron above the anode, and $z_{max}$ is the height of the field cage. The integral is truncated at $z_{max}/2$, i.e.\ the middle of the field cage in the drift direction, because the field distortions, and hence the transverse displacement, have the opposite sign above and below $z_{max}/2$. As a result, a particle drifting from the middle of the field cage to the bottom experiences the maximum displacement. We calculate this maximum transverse displacement using COMSOL~\cite{comsol}, a finite element analysis package, for various field cage designs. 

A consistent feature of all our field cage designs is rectangular rings, to fit within the vessel while maintaining 1~cm clearance with the vessel walls. This fixes the outer dimensions of the rings. The remaining parameters we can optimize are ring thickness, clearance between the inner edge of the rings and the sensitive volume, and the spacing between rings. We find that increasing the clearance and decreasing the spacing minimizes the average transverse displacement. We find no significant impact with different thicknesses. 

\begin{figure}
  \centering
  \includegraphics[width=1\columnwidth]{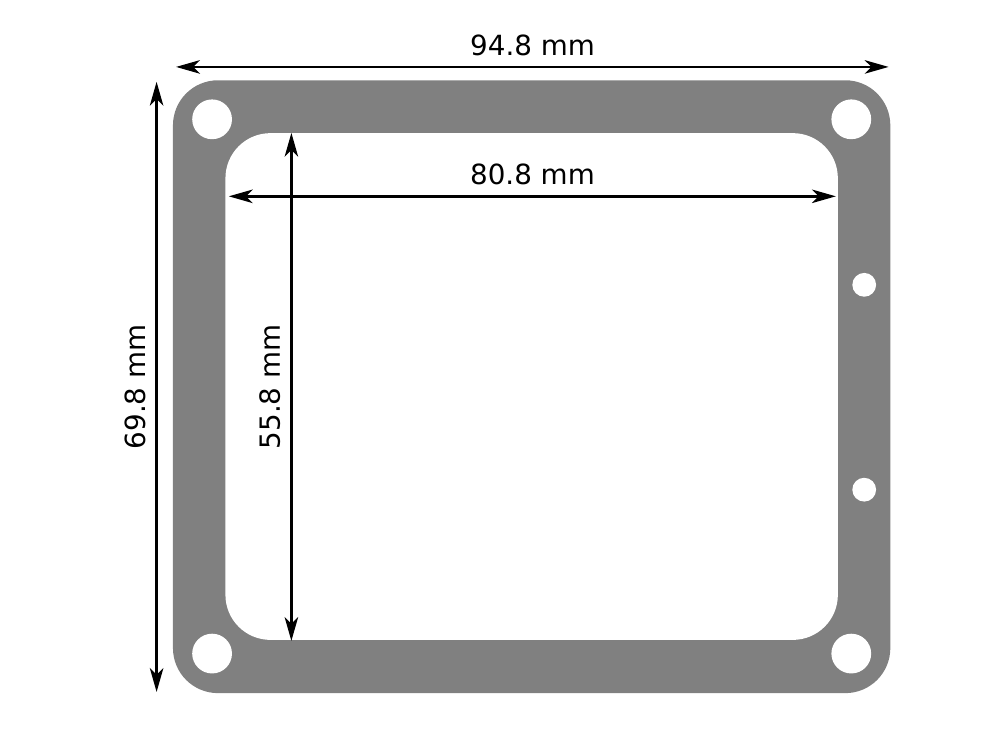}
  \caption{Optimized, rectangular field cage ring design used in the final BEAST TPC detectors.}
  \label{fig:recring}
\end{figure}

\begin{figure*}
  \centering
  \includegraphics[width=2\columnwidth]{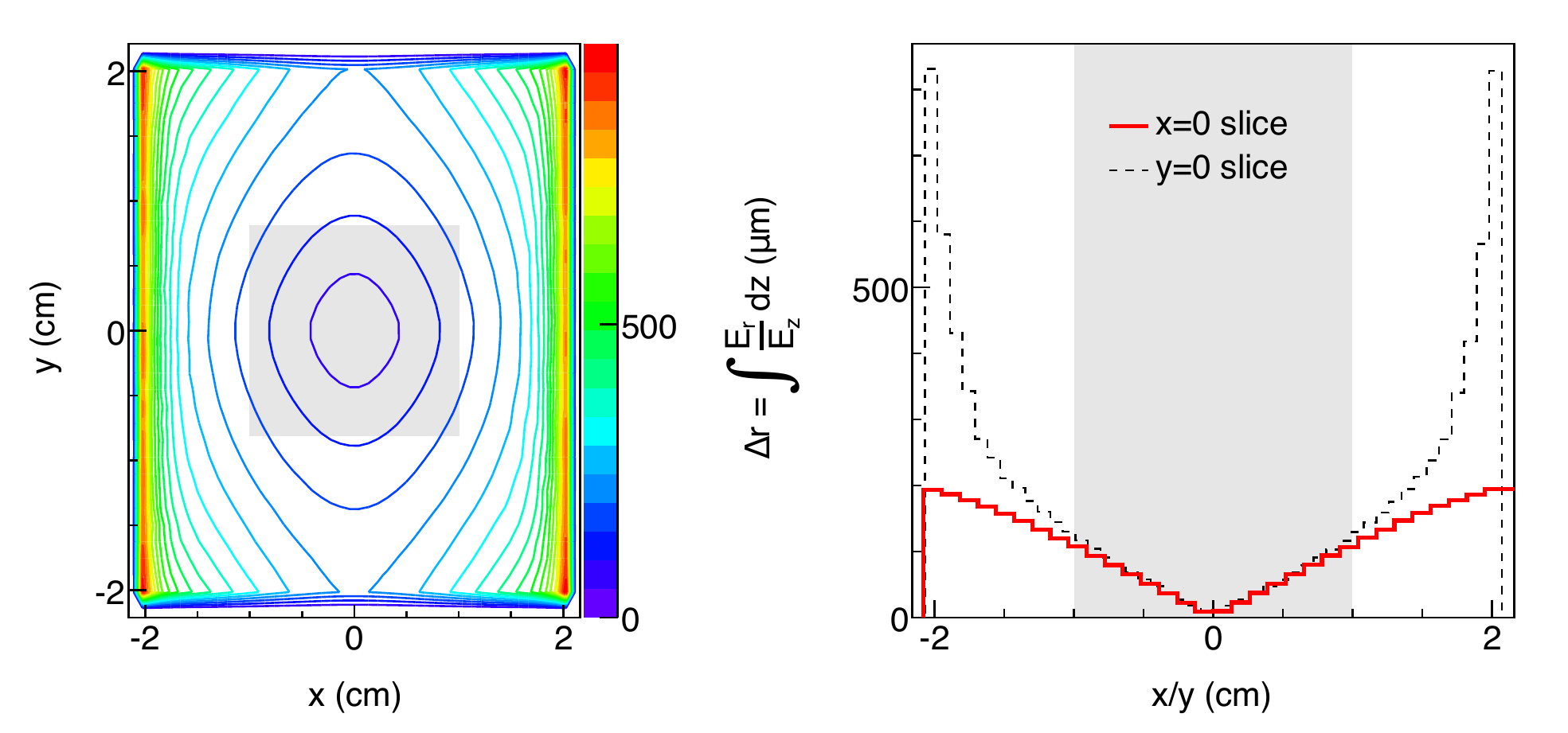}
  \caption{Simulated maximum transverse displacement $\Delta r(x,y)$ with the final, optimized field cage design. Left: $\Delta r$ versus position on the pixel chip plane $(x,y)$, defined so that $(x=0,y=0)$ is at the center of the chip. Right: $\Delta r$ for $x$=0 and $y$=0. In both figures the chip active area is shaded.}
  \label{fig:displacement}
\end{figure*}

We choose a field cage design (see Fig.~\ref{fig:recring}) that minimizes the ring spacing and maximizes the clearance between the rings and sensitive volume, within practical and mechanical limits. We therefore use rings with a thickness of 0.8~mm, with inner dimensions 7~mm inside the outer dimensions. We round the edges to minimize the risk of high voltage discharges. The approximate maximum transverse displacement for this design is shown in Fig.~\ref{fig:displacement}. We find that for this field cage design the maximum transverse displacement of ionization deposited above the active area of the pixel chip is approximately \SI{100}{\micro\meter}.

\subsubsection{Impact of external magnetic field}
We orient the TPCs so that the drift field is parallel or anti-parallel to the 1.5~T Belle II solenoid magnetic field. In order to study the impact of the magnetic field on the performance of the TPCs, we use the Magboltz~\cite{magboltz,Biagi:1999nwa} and Garfield++~\cite{garfield++} software packages to simulate from ideal (0 degrees) to worst-case (10 degrees) alignment scenarios. We simulate the drift of a large number of individual charges from a common starting point 10~cm above the anode and measure both their mean position and spread (RMS) after drifting in both fields.

\begin{figure*} 
  \begin{center}
    \includegraphics[width=2\columnwidth]{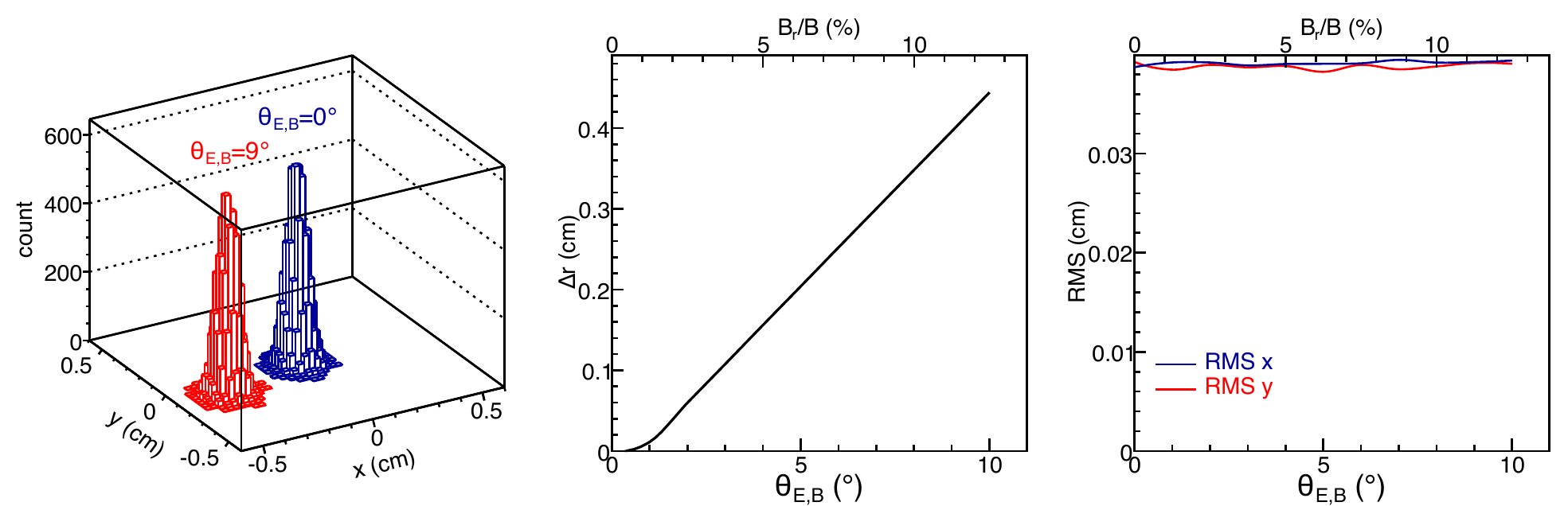}  
    \caption{Results from simulations of drifting charge in electric ($E$) and magnetic ($B$) fields. The electric field is parallel to the $z$ axis, as is the magnetic field, but with a small transverse component $B_r$ due to a misalignment angle, $\theta_{E,B}$, between the fields. Charges are placed in space at $(x,y,z)=(0,0,10~\text{cm})$ and allowed to drift to the $x-y$ plane at $z=0$. Left: a histogram of the final position of drifted charge for no misalignment and for a misalignment of 9 degrees. The effect of the misalignment is to displace the distribution on the $x-y$ plane. Middle: displacement on the $x-y$ plane versus misalignment. The displacement is taken as the distance between $(x,y)=(0,0)$ and the mean of the final position of the electrons after drift. Above 1 degree, the displacement is linear with misalignment. Right: spread (RMS) of the distribution of the final positions of the electrons after drift versus misalignment.}
    \label{fig:bfield}
  \end{center}
\end{figure*}

Our key findings are illustrated in Fig.~\ref{fig:bfield}. We observe that the presence of the magnetic field and its orientation have negligible effects on the spread of the charges after drift, consistent with our expectations that this is determined primarily by gas properties. We also find that the effect of a misaligned magnetic field is to displace the arrival position of the electrons by an amount proportional to the misalignment angle for misalignments above 1 degree. 

Based on this study we require alignment to be parallel to within 1 degree for Phase 2 operation (there is no solenoid in Phase 1). This ensures that the maximum displacement is below \SI{100}{\micro\meter}, i.e.\ of the same magnitude as the expected displacements due to drift field uniformity. 

\subsubsection{Gain uniformity and stability}
We do not expect the effective gain of the TPCs to be stable over time due to variations in gas conditions, pressure and temperature, and detector aging. To continuously monitor effective gain, we use the embedded $^{210}$Po alpha sources. The $5.3$~MeV alpha particles transect the sensitive volume, leaving a consistent and distinctive ionization trail. The variations in the recorded charge from these calibration tracks can then be used to monitor the effective gain variation with time, as presented in Section \ref{subsubsection:gain}, and to correct for such variations. In order to achieve optimal energy resolution for nuclear recoils, it is also important that the gain be spatially uniform, or that any non-uniformities are calibrated out. Our previous generation detectors in fact exhibited substantial effective gain variations due to non-uniform metallization of the pixel chip \cite{Vahsen:2014fba}. This issue has been resolved in the BEAST TPCs. Measurements of the uniformity with respect to position are presented in Section \ref{sec:dd_uniformity}.

\section{Description of built detectors}

We use both prototype and final TPCs in the first phase of SuperKEKB commissioning~\cite{Lewis:2018ayu}, and only final TPCs in later phases. Both types of TPCs are roughly shoebox-sized aluminum vessels with a single FE-I4B pixel chip, aluminum field cage, two GEMs, embedded alpha sources, and Parylene C coating as described in the previous section. In this section we will describe in more detail the specific features of both types of TPCs.

\subsection{Prototype pressure vessel}
The $11\times13\times31$\,cm$^3$ prototype pressure vessels consist of two parts: a rectangular aluminum box without a top (Fig.~\ref{fig:neutron_tpc}) and a separate aluminum lid. The vessel is sealed by compressing an o-ring between the lid and the vessel with screws through the corners of the lid. We found that these four screws are insufficient to prevent bowing of the lid, and observe small leaks when the prototype vessel is operated at positive gauge pressure. The lid of final pressure vessel has six screws, resolving this issue. Serial data and low voltage (LV) power, gas inlet and outlet, and HV all pass through both ends of the vessels via feedthroughs mounted to KF-16 or KF-40 flanges compressed onto the vessel wall with claw clamps. The holes are sealed with Viton o-rings with aluminum centering rings between the feedthroughs and the vessel. 

\subsection{Final pressure vessel}
\begin{figure}
\centering
\includegraphics[width=.8\columnwidth]{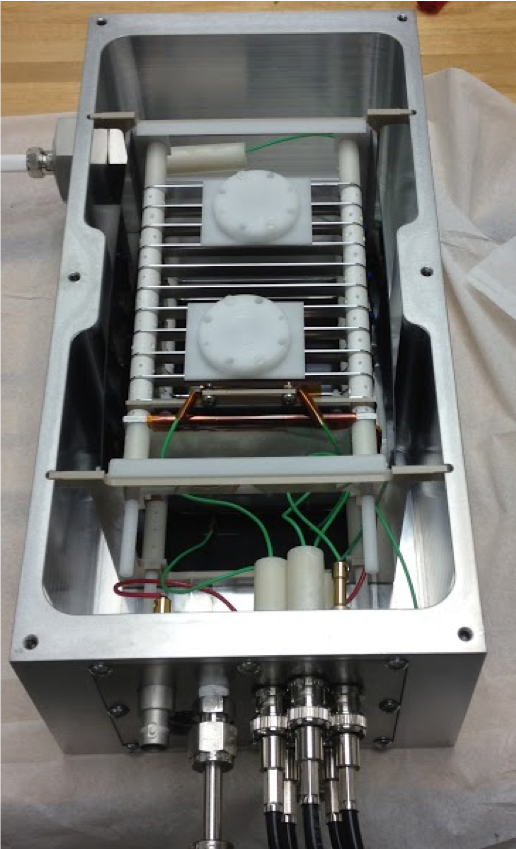}
\caption{Eight final BEAST TPCs were constructed and deployed. This photo shows the TPC pressure vessel without the lid but with the field cage installed. The feedthrough endplate with HV and gas services connected is visible at the bottom. The 10-kV HV feedthrough for the cathode is visible at the top left. The two white (acetal) disk-like structures contain the $^{210}$Po calibration sources.}
\label{fig:final_insides}
\end{figure}

The final pressure vessel measures $10~{\rm cm}~\times 15~{\rm cm}~\times 31~{\rm cm}$ and consists of three parts: a rectangular aluminum box without a top and with a large opening in one end (Fig.~\ref{fig:final_insides}), a separate aluminum lid, and a separate feedthrough endcap. To facilitate welding of feedthroughs onto the endcap, which was required to achieve sufficiently compact detectors, the endcap is manufactured out of Type 316L stainless steel. The endcaps are surveyed for magnetism after machining, and all have a magnetic permeability less than 1.4 times the permeability of vacuum. The three parts of the pressure vessel are sealed by o-rings located on the lid and feedthrough endplate using 6 and 8 screws, respectively. Serial data, LV power, gas inlet and outlet, and GEM HV feedthroughs are welded onto the removable feedthrough endcap plate. The field cage HV enters via a dedicated 10~kV-rated baseplate-style feedthrough mounted on the side of the vessel. To create stronger threads, tapped holes in the aluminum box have helical inserts made from INCONEL X-750, a nickel-chromium alloy \cite{x750}.

\subsection{Internal components}
\begin{figure*}
\centering
\includegraphics[width=1.6\columnwidth]{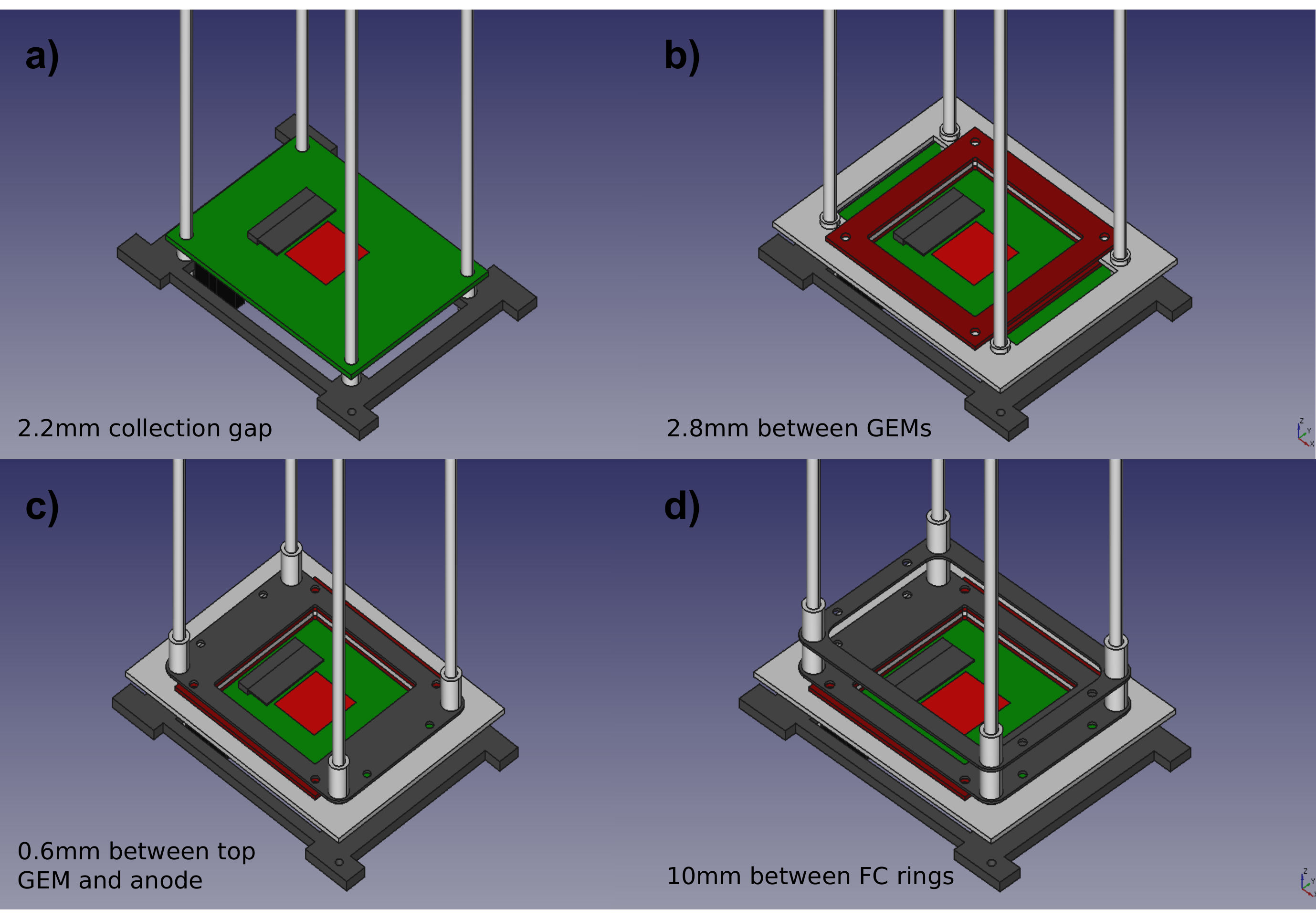}
\caption{(color online) A CAD rendering of the installation sequence for a TPC tower. Step a) shows the PEEK support plate (dark gray), Delrin threaded rods (white), and the pixel chip (red) mounted on its printed circuit board (green) with nearby wire bond shield (dark gray). Step b) adds the frames of the two GEMs (burgundy, with the GEM foils removed from the rendering for clarity) separated and supported by a PEEK spacer (white) to maintain a 2.8~mm transfer gap between the surfaces of the two GEM foils. There is a 2.2\,mm gap between the chip and the first GEM foil. Step c) adds the anode (dark gray), which is placed on the top GEM frame. Finally, step d) adds the first field cage ring (dark gray) with 1~cm Delrin spacers (white).}
\label{fig:assembly}
\end{figure*}

\begin{figure}   
\centering       
\includegraphics[width=\columnwidth]{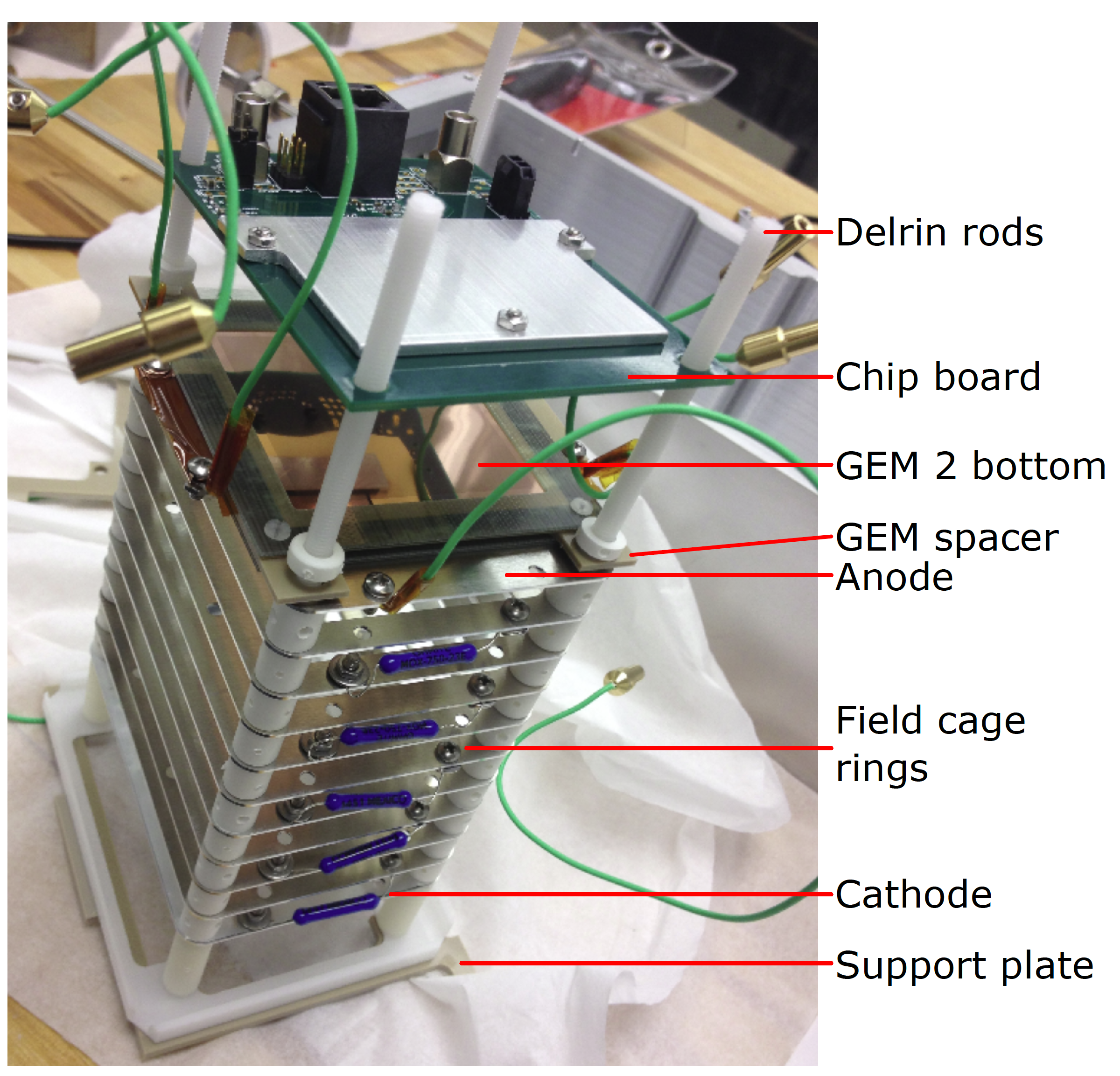}
\caption{One TPC tower during assembly, shown inverted compared to Fig.~\ref{fig:assembly}. For clarity, we have separated the pixel chip board from the GEMs. The second support plate is not shown and will be attached after the chip board is secured.}
\label{fig:tower}
\end{figure}

The key internal components of the TPCs are the pixel chip, GEMs, and field cage. Other internal components are auxiliary to these, providing structural support and electrical services. We assemble these into an integrated structure called a ``tower''  outside the vessel in order to simplify the assembly and testing. One tower consists of four parallel Delrin (acetal) threaded rods held in place by two PEEK (aluminum in prototype) support plates, one at the bottom and the other at the top, with all other components sandwiched between these two support plates, separated with Delrin spacers. Fig.~\ref{fig:assembly} shows the assembly sequence of the first layers of a tower, and Fig.~\ref{fig:tower} shows a final TPC tower during assembly. 

We assemble and test each tower in a class 10,000 clean room. After assembly and testing is complete we place the tower in a vessel, with the support plates sliding into a pair of slots milled into the inner walls of the vessel (see Fig.~\ref{fig:final_insides}).

\subsection{High voltage}
\begin{figure}
\centering
\includegraphics[width=0.8\columnwidth]{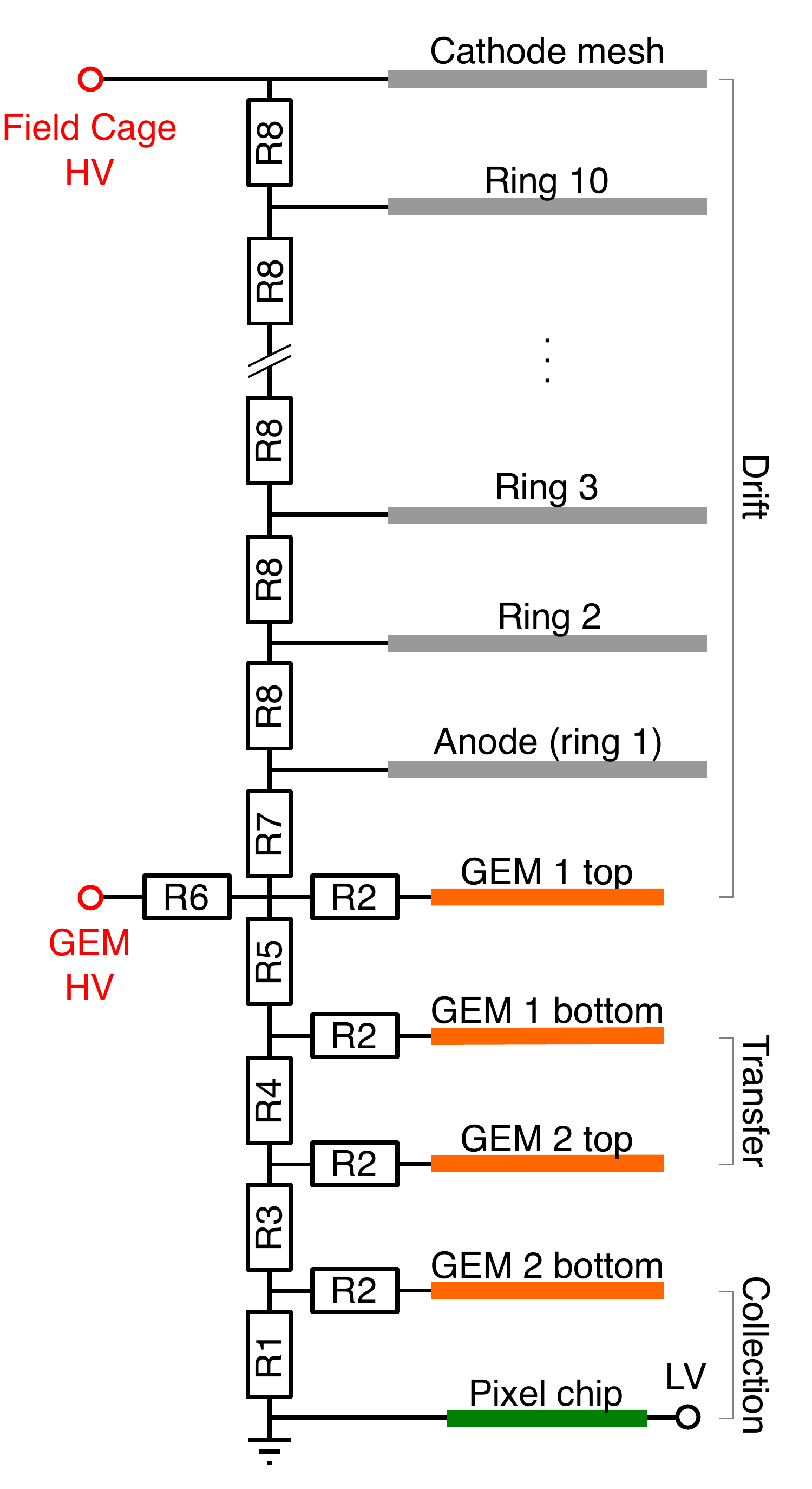}
\caption{A schematic of the HV circuit inside one TPC. Resistance values are chosen in order to set the optimized drift field (R8, R7), transfer field (R4), collection field (R1), and gain (R3, R5). Other resistors are for current limiting.}
\label{fig:hv_schematic}
\end{figure}

The HV circuits in the TPCs provide the drift field (inside the field cage), transfer field (between GEMs), collection field (between the bottom GEM and pixel chip) and amplification voltage (across GEMs). In the prototype TPCs, we use potentiometers located outside the vessels in place of some resistors for flexibility. In the final TPCs, all resistors are located inside the vessels and the resistance values are fixed to achieve the optimal fields and gain (see Fig.~\ref{fig:hv_schematic}). 

Each TPC utilizes two HV channels, which allows for independent adjustment of the drift field and gain: the ``double GEM voltage'' sets the total voltage across both GEMs plus the voltage across the transfer and collection gaps. The ``field cage voltage'' provides the drift field. The field cage and amplification circuits are connected, so the voltage across the field cage is equal to the field cage voltage minus the GEM voltage. Table~\ref{tab:resistors} summarizes the voltages, resistances and fields between each layer.

\begin{table*}
\caption{Summary of the properties of the resistor ladder in the final TPC HV system, following Fig.~\ref{fig:hv_schematic}. Each row in this table corresponds to one resistor in the ladder, which corresponds to a specific gap and field. The first column contains the layer label, the second the resistor label, the third the resistance $R$, the fourth the voltage drop $\Delta V$ across the resistor, the fifth the length $\Delta z$ of the volume associated with the layer, and the last the resultant electric field in this volume. Here we have assumed that the field cage voltage is $-8$~kV and the GEM voltage is $-2.1$~kV, which correspond to the ideal drift velocity and gain. The current-limiting resistors outside the ladder have values R2, R6=\SI{300}{\mega\ohm}.} \label{tab:v_r_E}
\begin{center}
\begin{tabular}{lccccc}
        \toprule
        Layer                          & Resistor & $R$~[M$\Omega$]  & $\Delta V$~[V]   & $\Delta z$~[cm]   &  $E$~[V/cm]   \\
        \midrule
        Collection gap                 & R1       & 11               & 700          & 0.22         & 3170  \\
        GEM 2                          & R3       & 6                & 380          & 0.005        & 76400 \\
        Transfer gap                   & R4       & 10               & 640          & 0.28         & 2260  \\
        GEM 1                          & R5       & 6                & 380          & 0.005        & 76400 \\
        Drift gap (GEM 1 top to anode) & R7       & 6                & 33         & 0.06         & 544   \\                                               
        Drift gap (between two rings)  & R8       & 100              & 540          & 1.00         & 544   \\
        Drift gap (anode to cathode)   &          & 1000             & 5900           & 10.87        & 544   \\
        \bottomrule
\end{tabular}
\end{center}
\label{tab:resistors}
\end{table*}

\subsection{Gas system}
The BEAST TPCs are designed to operate with a small flow (${\sim}10$~sccm) of premixed He:CO$_2$ at a 70:30 ratio. We use a mass flow controller to maintain a steady flow of gas, with additional pressure and flow gauges. In lab tests we pump air out of the vessels before purging with the target gas, but in BEAST operation we simply flow (at a maximum rate of 250~sccm) until achieving gain.

\section{Charge readout and data acquisition}\label{sec:daq}

The \mbox{FE-I4B} pixel readout chip was originally designed for use in the pixel vertex detector of the \mbox{ATLAS} experiment at the Large Hadron Collider~\cite{atlas_collaboration_2008}. There, one is typically interested in assigning detected charge pulses to particular bunch crossings (BC) of the accelerator beams, which occur every 25\,ns~(equivalent to 40\,MHz). To allow for timing-in of the pixel detector, readout of up to 16 consecutive BCs was foreseen by the \mbox{ATLAS} collaboration. However, to read out large ionization clouds with the \mbox{FE-I4B} in a TPC, it is necessary to detect charge pulses in a much larger number of consecutive \mbox{25-ns} periods. The readout of up to 255 consecutive time periods is achieved by implementing a custom readout sequence in the data acquisition (DAQ) system. In this section we describe the design and operation of the chip, to provide context, and then the implementation of this custom readout sequence.

\subsection{FE-I4 pixel readout chip}

The \mbox{FE-I4}-type pixel readout chip has been developed for future ATLAS pixel detector upgrades, in particular the ATLAS Insertable B-layer upgrade~\cite{ibl_collaboration_2012,ibl_collaboration_2018}. The chip is manufactured in an IBM 130\,nm bulk CMOS process. It has an area of $20.2\times18.8$\,mm$^2$ and comprises 26,880 pixel cells arranged in 80~columns and 336~rows, each pixel measuring $250\times50$\,\SI{}{\micro\meter^2}. 

The analog part of each pixel holds a two-stage charge sensitive amplifier~(CSA), optimized for low power consumption, low noise and fast rise time. The amplification stage is followed by a comparator, which compares the amplifier signal to a threshold voltage. The comparator outputs logical high when the amplifier signal is above the threshold voltage. The measured time interval the comparator signal is high~(TOT, short for Time-over-Threshold) is to first order proportional to the charge collected by the CSA. The achievable charge resolution depends on the clock frequency used for measuring the TOT (nominally 40\,MHz), the fall time of the CSA (determined by the feedback current of the CSA; typically a few clock cycles), and the number of storage bits for the TOT value in the \mbox{FE-I4}~(dynamic range of four bits). Each analog pixel is configured through global registers affecting all pixels~(e.g., bias currents, global feedback currents and global comparator threshold voltage) and pixel registers affecting individual pixels~(e.g., local feedback current and local comparator threshold voltage), which superimpose on top of the global register settings.

A single digital region processes the information coming from $2 \times 2$ pixels~(\mbox{four-pixel} digital region). Five latency calculation and triggering units are shared by the four pixels.
Each pixel has five storage cells for TOT information, each assigned to a latency calculation and triggering unit. The time stamp, which is stored in the latency calculation and triggering unit, is common to all simultaneous hits but the TOT information is calculated and stored individually for each pixel. The trigger latency is the duration over which the TOT information is preserved in memory and can be adjusted via a global register from one to 255 clock cycles at 40\,MHz. The TOT information is stored locally and only transferred to the chip periphery for further processing when a trigger is issued.

The latency calculation and triggering unit has a counter that counts down from 255 to zero when a new hit is detected. The hit data (i.e., TOT information, column and row addresses) are made available for readout when the trigger signal arrives exactly when the counter has fallen below the trigger latency. A missing trigger signal resets the latency calculation and triggering unit and permits processing of a new hit. The \mbox{FE-I4} allows up to 16 consecutive triggers~(equivalent to 400\,ns), distributed globally to all latency calculation and triggering units. The hit data are read from the local buffers, reorganized and 8b/10b encoded in the chip periphery, and then transferred serially at 160\,Mbit/s for data acquisition.

\subsection{Data acquisition}

Each \mbox{FE-I4} chip inside a vessel is mounted on a custom interface board that routes power, LVDS communications, and trigger lines between the chip and the readout system, located outside the vessel. The interface board includes a grounded aluminum bridge over the wire bonds for protection. For the generation of a trigger signal, individual pixels can be selected via the pixel register and contribute to the so-called \mbox{HitOR} signal. The \mbox{HitOR} signal, which is available on a pad of the \mbox{FE-I4} pixel readout chip, is the logically OR’ed comparator output of all selected pixels and works independently from digital processing of the hit data. The single-ended \mbox{HitOR} signal is converted inside the vessel to a differential signal, which is transmitted via twisted-pair cable up to 40\,m to the readout system. The HitOR signal triggers the readout system, which generates the readout sequence to retrieve the hit data from the detector.

\subsubsection{Readout system}

The pyBAR readout software~\cite{pybar} is a versatile readout and test system for \mbox{FE-I4}-type pixel readout chips. The software was purposely built to investigate chip tuning methods~\cite{janssen_2017} and to accommodate the research and development of new sensor technologies for hybrid pixel detectors~\cite{pohl_2017}.
The pyBAR software is designed to support different hardware platforms that can be selected depending on the application.
Areas of application range from single-chip readout for sensor characterization~\cite{pohl_2015} to multi-chip readout for operation and testing of larger-scale detectors~\cite{gonella_2017, Lewis:2014poa}. PyBAR and the underlying data acquisition framework Basil~\cite{basil} are written in Python and provide FPGA firmware and firmware modules written in Verilog. The readout system enables fast tuning of the FE-I4 chips and provides online-monitoring software for displaying hit data in real time.

Three readout platforms supported by pyBAR were used to read out the BEAST TPCs at different stages of the project. During production and in Phase 1, we used SEABAS2 and \mbox{USBpix2}. SEABAS2 is a custom-built readout board containing a Xilinx Virtex-5 FPGA interfacing with the FE-I4 chips. A network stack~\cite{uchida_2008} is implemented in a Xilinx Virtex-4 FPGA for Gigabit Ethernet communication with a DAQ computer. A daughter board attached to SEABAS2 allows the connection of up to four FE-I4 chips. USBpix2 is a custom-built multi-purpose readout board containing a Xilinx Spartan-3 FPGA and a Cypress \mbox{EZ-USB} FX2 USB~2.0 microcontroller. Two different adapter cards are available for operating a single \mbox{FE-I4} chip and up to four \mbox{FE-I4} chips. In Phase 3, we used MMC3, a custom-built base board for readout of up to 16 \mbox{FE-I4} chips. The base board is designed to carry a commercially available FPGA board containing a Xilinx Kintex-7 FPGA. The FPGA interfaces with the FE-I4 chips and includes an implementation of a network stack~\cite{uchida_2008} for Gigabit Ethernet communication with a DAQ computer. The pyBAR firmware for Phase~2 supports up to 8 \mbox{FE-I4} chips, which can be read out simultaneously and triggered independently. All platforms allow continuous and simultaneous readout of several FE-I4 chips at a trigger rate of 200\,kHz.

\subsubsection{Readout sequence}

\begin{figure}[ht]
  \centering
  \includegraphics[width=\columnwidth]{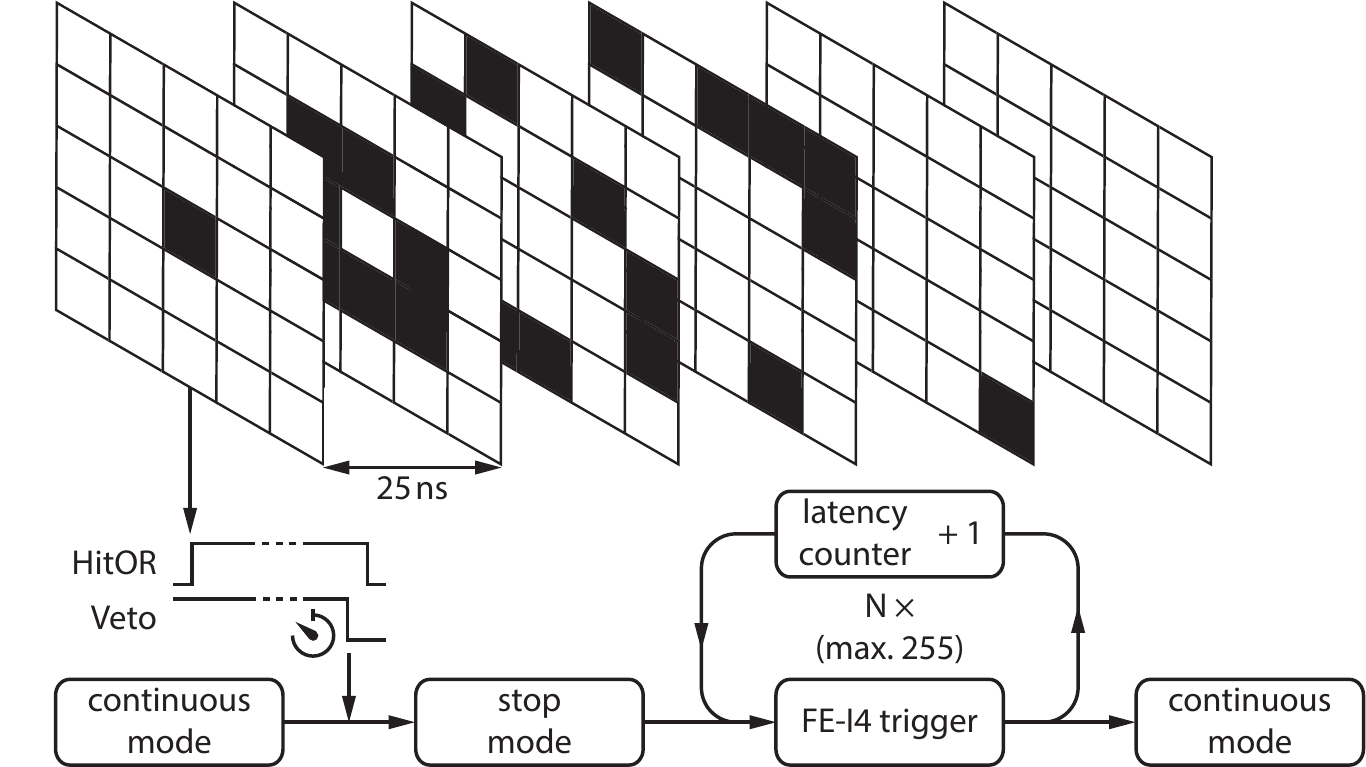}
  \caption{Schematic representation of the custom pixel chip readout sequence that allows data from up to 255 consecutive BCs~(equivalent to \SI{6.4}{\micro\second}) to be recorded.}
  \label{fig:stop_mode}
\end{figure}

In pyBAR a readout sequence using the \mbox{FE-I4} stop mode has been implemented, which allows the readout of up to 255 consecutive BCs (equivalent to \SI{6.4}{\micro\second}, see Fig.~\ref{fig:stop_mode}). In stop mode, the clock to the pixel matrix is stopped, the latency counters are frozen for all pending hits, and no new hits will be stored. This preserves the hit information and allows the readout of the hit information at any time. In order to receive the hit information, sending a single trigger signal and advancing the latency counters by one must be alternated. This can be repeated up to 255 times until the counters overflow and no stored hits remain. Once the readout sequence has been completed, the \mbox{FE-I4} can be brought back into continuous mode to store new hits. The stop mode readout also involves the HitOR signal, which is passed to and processed inside the FPGA. Whenever a hit is detected, the HitOR goes high and triggers a command sequencer module inside the FPGA that sends a command sequence to the \mbox{FE-I4}. The command sequence is generated on the FPGA and can have a length of up to several hundred thousand bits, which are sent to the \mbox{FE-I4} at a rate of 40\,Mbit/s. The trigger rate is then limited by the length of the command sequence, i.e. the number of consecutive BCs designated for readout, and the time required to retrieve all hit data. A trigger rate on the order of 10\,Hz can be achieved when retrieving data from 255 consecutive BCs with several thousand hits. Higher trigger rates can be achieved by reducing the number of consecutive BCs to be read.

Without this readout sequence using the stop mode, no tracks with longer than 4\,mm extent in the $z$-direction could be recovered, which includes not only a large portion of the nuclear recoils, but also the tracks with the best directional determination. The readout sequence was used with either 255~(Phase~1) or 100~(Phase~2) cycles.

\subsubsection{Trigger veto}

The length of a HitOR signal encodes information about the charge contained in a track and the extent of the track in z-direction. Electron recoils caused by X-rays interacting with the target gas are extremely copious in realistic operating environments. HitOR pulses generated by these electron recoils are typically very short compared to those generated by nuclear recoils. Therefore, HitOR signals are rejected at the firmware level and a readout sequence is only issued to the \mbox{FE-I4} chip if the HitOR length is sufficiently long (typically above 400\,ns) so that the track is unlikely to be an electron recoil. The trigger module of the pyBAR FPGA firmware allows variable adjustment of the required signal length in steps of 25\,ns.

\section{Testing and performance}

\label{sec:performance}

All TPCs undergo a series of quality control (QC) tests to validate their construction and basic performance. We describe here the purpose and results of these tests.

\subsection{Quality control and acceptance criteria}
Our QC procedure consists of a number of production tests that each TPC undergoes.  In addition, as part of the U.S. Department of Energy Belle II construction project, formal criteria for minimal acceptable ({\it threshold}), desired ({\it objective}), and design-goal ({\it  specification}) performance were established, see Table \ref{tab:criteria}. These criteria flow from the desire to measure the directional distribution and energy spectrum of neutron recoils, with recoil energies of order 100\,keV to a few MeV, in multiple high-background locations in Belle II. Because of the scattering angle involved in the elastic scattering, the angular resolution objective is modest. The gain, gain stability, and energy resolution objectives are chosen to enable measurements of the exponentially falling recoil-energy spectrum expected at SuperKEKB.

\begin{table*}
\caption{Summary of formal TPC acceptance criteria and achieved performance. All performance threshold values are satisfied for all TPCs. For most parameters, the measured performance exceeds the specification and significantly exceeds the objective. For the gain, however, we found that a lower gain, near 1500, was in fact optimal, and hence only tested the first few detectors at higher gains.}

\begin{center}
\begin{tabular}{lcccc}
        \toprule
         Quantity                          					& Threshold & Objective & Specification & Achieved\\
         \midrule
	Angular resolution (1-cm tracks	)			&	n/a		& 	15\degree		& 5\degree & 2.5\degree	 \\
	Gain									&	1,000	& 10,000 			& 20,000	& 50,000\\	
	Gain stability, one week					& 	n/a 		& 20\%			& 5 \%	  & 1\% \\
	Energy resolution at 5.9 keV				&	n/a		& 20\% 			& 12 \%    &10\%\\
        \bottomrule
\end{tabular}
\end{center}
\label{tab:criteria}
\end{table*}

\subsection{Gas quality and leak checking}
The first QC test validates detector cleanliness and absence of leaks. We start by pumping down, and when the pressure is between 50 and 100~mTorr we flush the vessel three times in a row for a second, with the flow controller set to maximum, and continue to pump down for three hours. Then, we stop the pump and measure the pressure for at least two hours, looking for an increase in pressure due to outgassing. In all TPCs we observe no leaks at negative gauge pressure and find that the outgassing curves do not show any unacceptable contamination.

To check for leaks at positive gauge pressure, we rapidly fill the vessel with 800~Torr of He:CO$_2$ and monitor the pressure for fifty minutes. In these tests, all TPCs come to an equilibrium pressure above atmospheric pressure and we therefore conclude that there are no significant leaks.

\subsection{\label{subsubsection:gain}Gain, energy resolution, and stability}
Each TPC must satisfy three objectives related to effective gain: that they can achieve a minimum effective gain of 10$^4$, that the energy resolution (which is dominated by the gain resolution) at 5.9~keV is better than $20\%$, and that the gain is stable to better than $20\%$ over long and short timescales.

\begin{figure}
\centering
\includegraphics[width=\columnwidth]{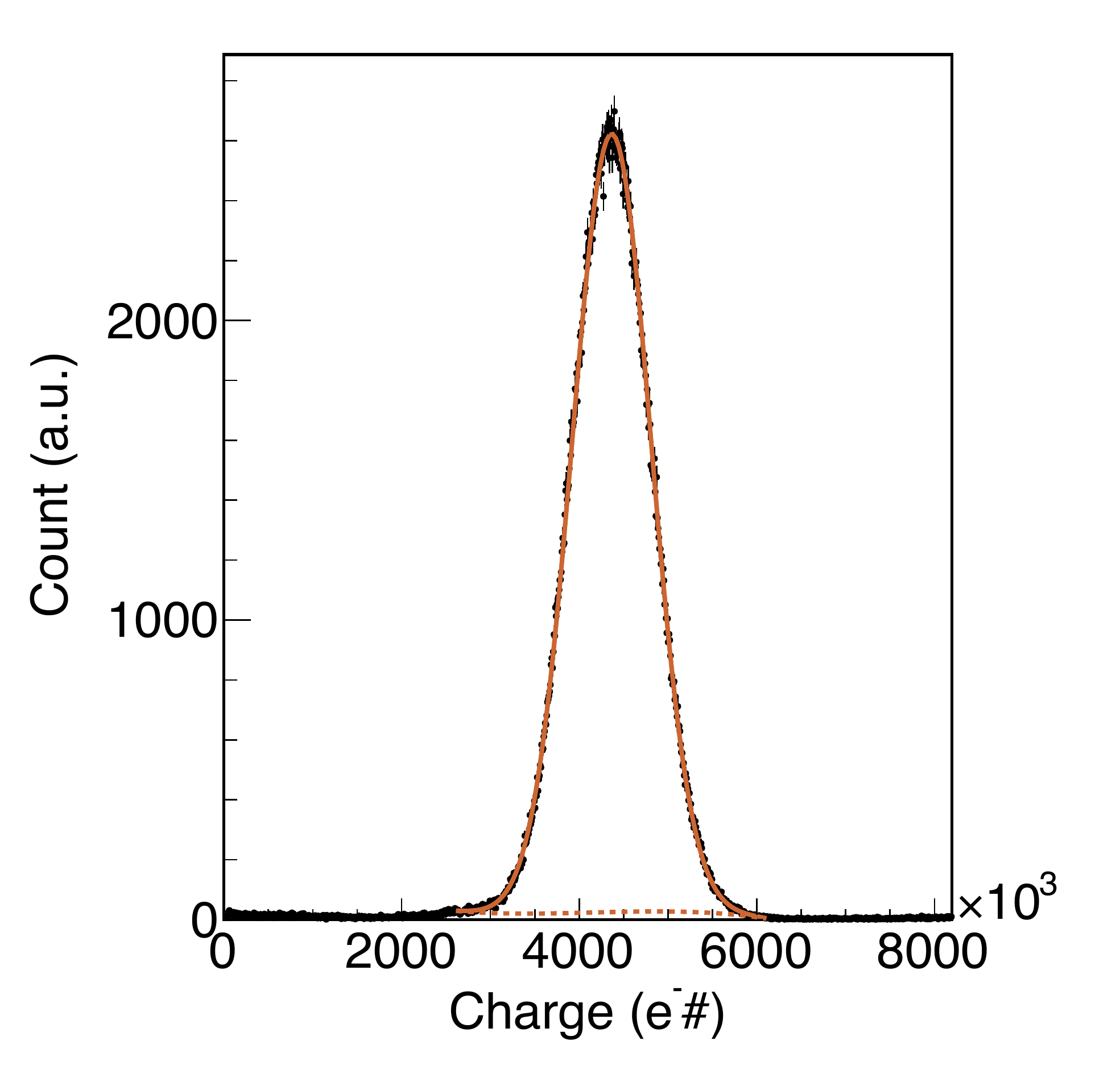}
\caption{(color online) Amplified charge measured by the chip of a final TPC during testing with an $^{55}$Fe X-ray source. The black points show experimental data, while the lines show the result of fitting the data with a third-order polynomial for the background (dashed red) and Gaussian distribution for the signal. The solid red line is the sum of the fitted signal and background.}
\label{fig:effective_gain_chip}
\end{figure}

To address the minimum effective gain and its resolution, we place an uncollimated $^{55}$Fe source on the cathode mesh with the source opening facing the chip. In He:CO$_2$ we expect to see only the 5.9~keV photon conversions, which should produce roughly 168 electron-ion pairs (estimated with Garfield++~\cite{garfield++}), mostly near the mesh. We read the amplified charge from a pad of copper that surrounds the pixel chip and also with the chip itself. The copper pad is connected to a pulse-height analyzer (PHA) as in Ref.~\cite{Vahsen:2014fba}. When the pixel chip is used, the avalanche charge of each event is measured by converting the TOT values of all pixels with hits into charge values, also as in Ref.~\cite{Vahsen:2014fba}. A measurement of the avalanche charge, $Q$, from an ionization event in the sensitive volume is essentially equivalent to measuring the primary ionization energy, $E$, because these quantities are related via

\begin{equation}
Q=\frac{EG}{W},
\end{equation}
where $G$ is the gain and $W$ is the ionization energy per electron-ion pair for the fill gas. We measure the avalanche charge distributions for different GEM voltages. Figure~\ref{fig:effective_gain_chip} shows the charge distribution measured by the chip. We extract the gain from a fit to this distribution, with a Gaussian for the signal and a third-order polynomial for the background. The effective gain is then equal to the peak position of the Gaussian divided by 168.

\begin{figure}
\centering
\includegraphics[width=\columnwidth]{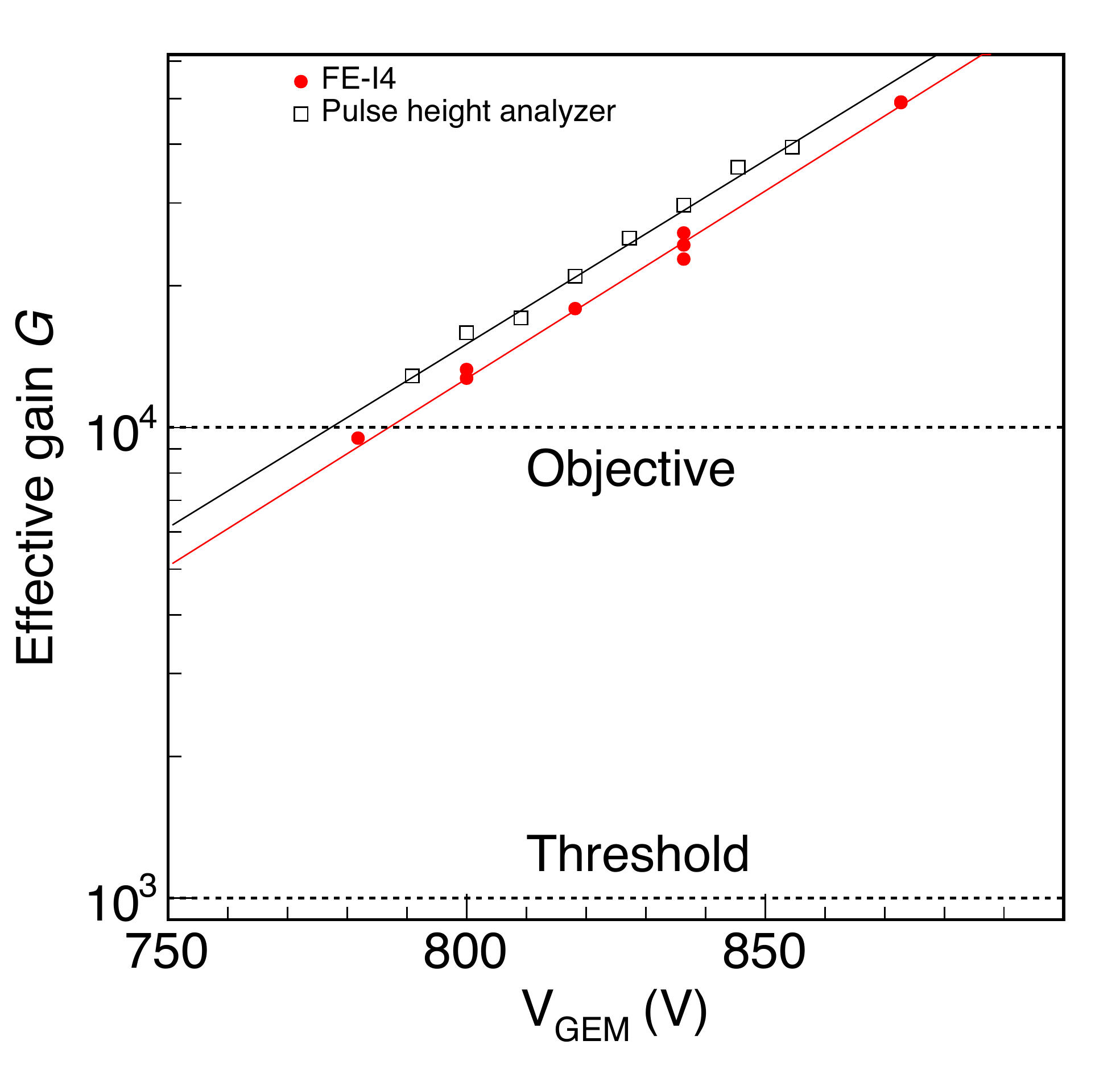}
\caption{Effective gain, $G$, as function of the double-GEM voltage for a final TPC. Red points: chip measurements. Black open squares: copper pad measurements. The two dashed lines show the objective and threshold gain values.}
\label{fig:effective_gain_v}
\end{figure}

Figure~\ref{fig:effective_gain_v} shows the effective gain as a function of the GEM voltage for both the PHA and pixel chip readouts. The two methods of measuring the effective gain agree to within $20\%$. All TPCs tested achieved well over the minimum effective gain required by the objective. 

\begin{figure}
\centering
\includegraphics[width=\columnwidth]{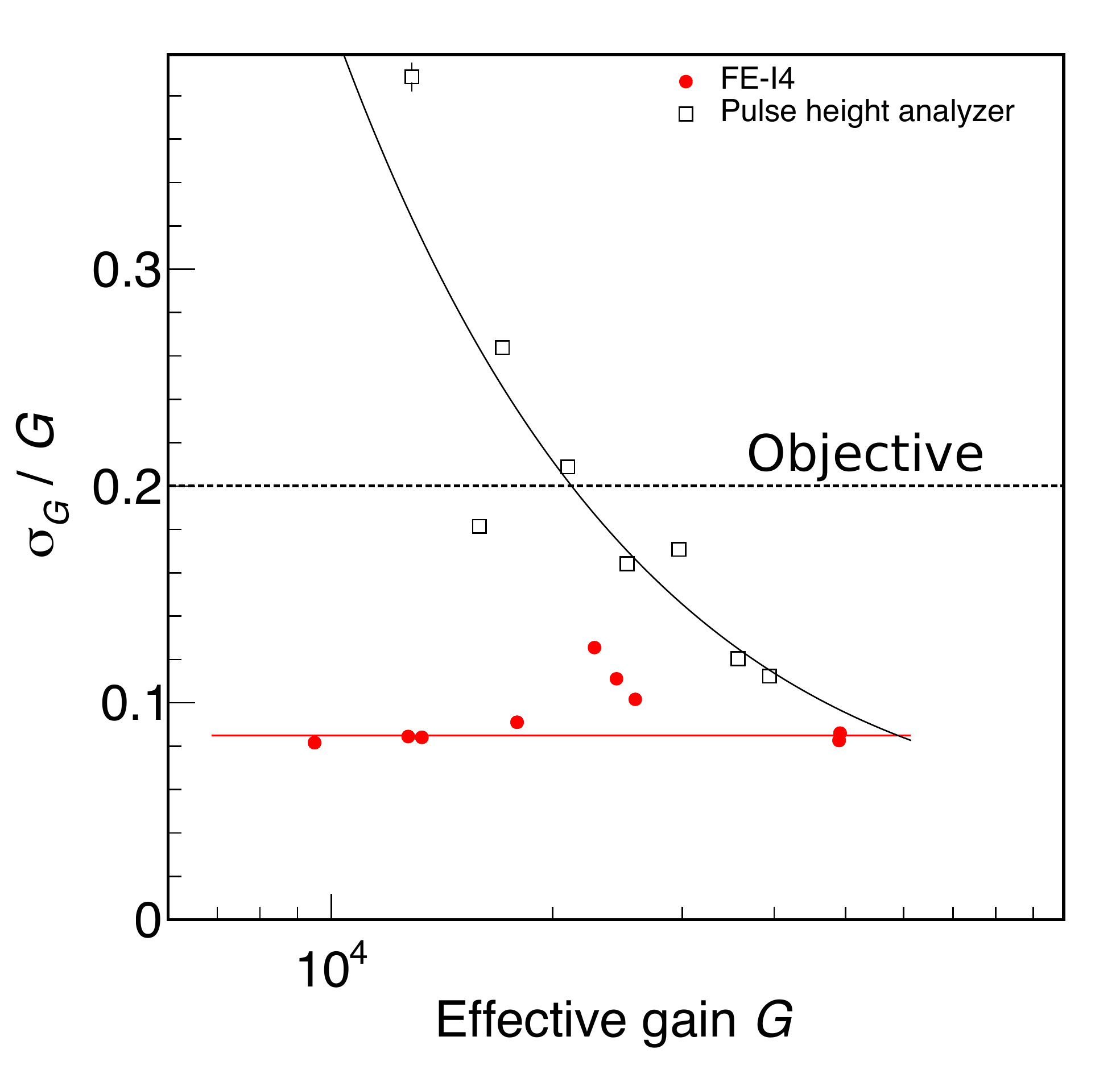}
\caption{Gain and energy resolution as function of the effective gain. Red point: chip ionization energy measurement. Black open square: copper pad gain measurement.}
\label{fig:gain_resolution}
\end{figure}

To measure the gain resolution and energy resolution, we use the same dataset and fits as above. We define the gain resolution as the ratio of the Gaussian width $\sigma_{\textrm{G}}$ in the fit to the charge distribution to its mean position. This gain resolution will include fluctuations in primary ionization statistics, statistical and spatial variations in GEM gain, and noise in the PHA system. We define the energy resolution as the same ratio of $\sigma_{\textrm{G}}/G$, but this time taken from the fit to the charge distribution from the pixel chip. This energy resolution will include all the aforementioned fluctuations, except for noise in the PHA, and in addition include contributions from non-uniformity in the pixel chip response and calibration, and contributions from the finite resolution of the charge measurement via TOT. Figure~\ref{fig:gain_resolution} shows the gain resolution and energy resolution as function of the effective gain.  As predicted in Ref.~\cite{Vahsen:2014fba}, the energy resolution in the pixel chip is much better than the apparent gain resolution in the copper pad for low gains. We attribute this to the much lower noise floor of the chip, which is determined by the capacitance and analog front end of individual pixels, and is of the order of 100 electrons. The PHA noise floor on the other hand is determined by the capacitance of the macroscopic copper pad used to collect the charge, and the downstream PHA signal chain. We find that with the chip the energy resolution is roughly 10$\%$ across a wide range of effective gains for all TPCs, satisfying the energy resolution requirement. The fact that the energy resolution is close to the asymptotic gain resolution at high gain also demonstrates that the four-bit TOT resolution per pixel (see Section~\ref{sec:daq}) and any non-uniformity related to the pixel chip are sub-dominant and negligible contributions to the final energy resolution.

\begin{figure}
\centering
\includegraphics[width=\columnwidth]{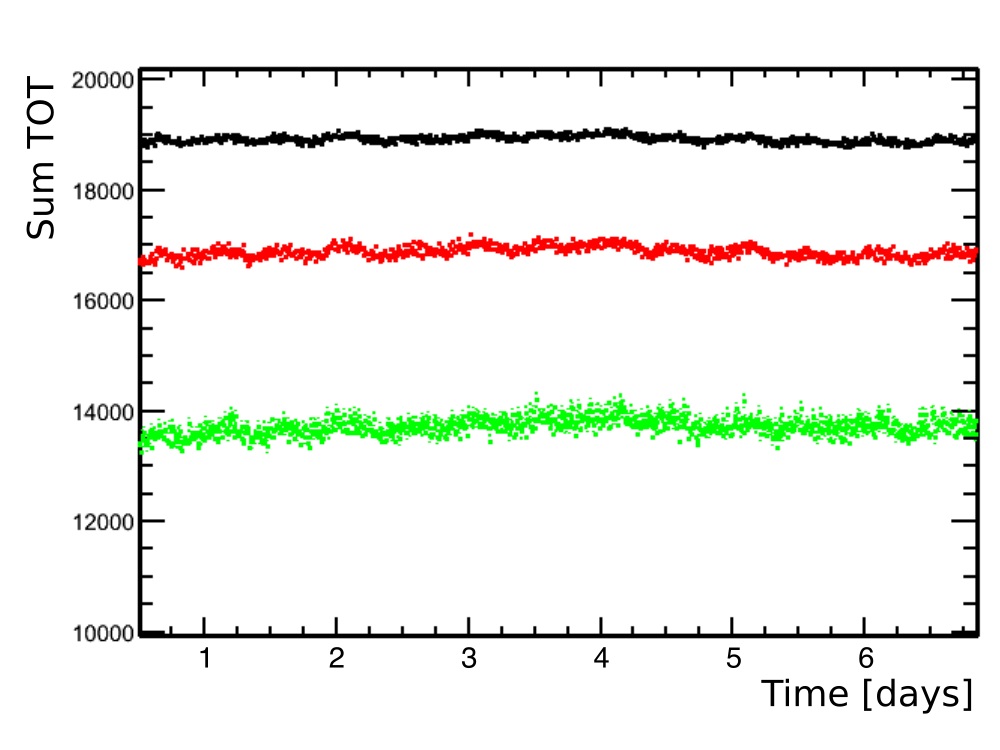}
\caption{Effective gain as measured by the summed charge (in TOT units) of tracks from three embedded $^{210}$Po alpha sources during a 1-week run. Each data point is an average over ten minutes. The effective gain differs between the sources due to diffusion and charge recombination. However, the effective gain of each source remains constant to within $1\%$ throughout the week.}
\label{fig:gain_stability}
\end{figure}

To measure gain stability, we monitor the total charge collected by the pixel chip from alpha tracks from the embedded $^{210}$Po sources over time. Figure~\ref{fig:gain_stability} shows the average collected charge (in TOT units) for three alpha sources at three different $z$ positions for a one-week period. We find that the effective gain of all three sources remains constant within $1\%$ (RMS) throughout the week, substantially exceeding the gain stability objective of 20\%.

\subsection{Tracking performance}
In order to achieve the physics goals of the BEAST TPCs, we require a recoil angle measurement precision of 15 degrees for 1-cm tracks. To encapsulate the effects of drift, amplification, and digitization, we measure the variation in the measured angle and charge of a standardized ionization track over repeated measurements. For the standardized tracks, we use the embedded $^{210}$Po sources and select alpha tracks in a narrow range of angles. Alpha tracks in our sensitive volume manifest as high-charge density, straight tracks (see Fig.~\ref{fig:event_displays}) that cross at least two sides of the chip and are therefore trivial to identify. 

\begin{figure}
\centering
\includegraphics[width=\columnwidth]{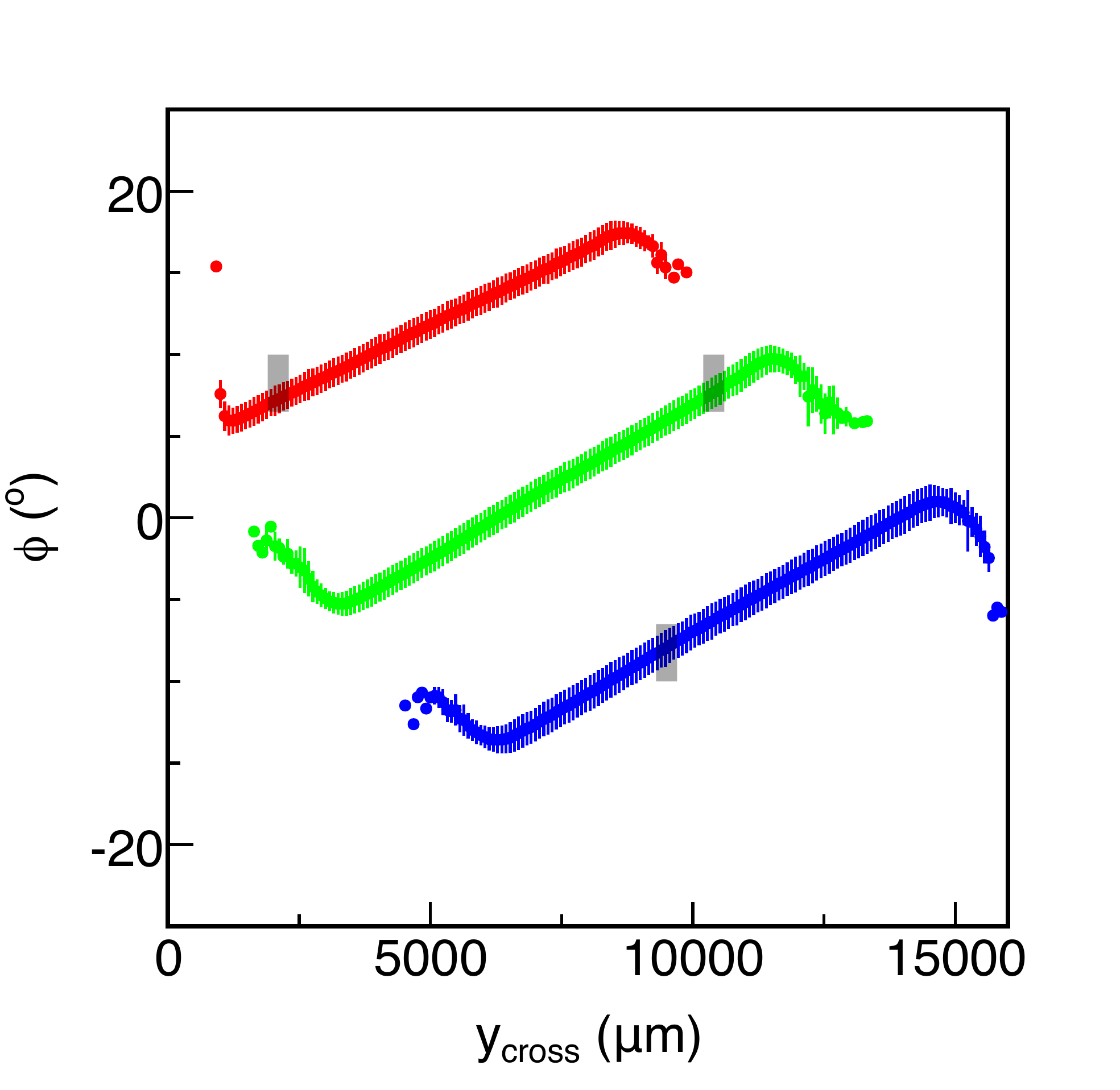}
\caption{Calibration alpha track $\phi$ vs. $y_{\textrm{cross}}$ ($y$ position where the track crosses the edge of the chip closest to the source) for a prototype TPC. The top source (red) is offset from the middle (green) and bottom (blue) sources due to mounting the sources at different $y$ positions. The shaded boxes indicate unique samples of tracks from each source with the same angle $|\phi|$, constituting the standardized alpha track samples.}
\label{fig:source_phi_vs_impact_parameter}
\end{figure}

In the following precision studies we use a sample of alpha tracks from each of three sources in a single prototype TPC. We consider only tracks within a narrow range of elevations with respect to the chip, between 9 and 17 degrees.  The sources are mounted not only at different $z$ positions but also at different $y$ positions, therefore we can uniquely identify the source of each alpha track based on its position in $\phi$ vs. $y_{\textrm{cross}}$ space. Fig.~\ref{fig:source_phi_vs_impact_parameter} shows distributions of these two variables and the selections we use to define the standardized alpha track collections: we define $y_{\textrm{cross}}$ windows of width \SI{400}{\micro\meter} so that the mean absolute value of $\phi$ of the tracks in the window is identical for all three windows. 

\subsubsection{Angle measurement precision}
\begin{figure}
\centering
\includegraphics[width=\columnwidth]{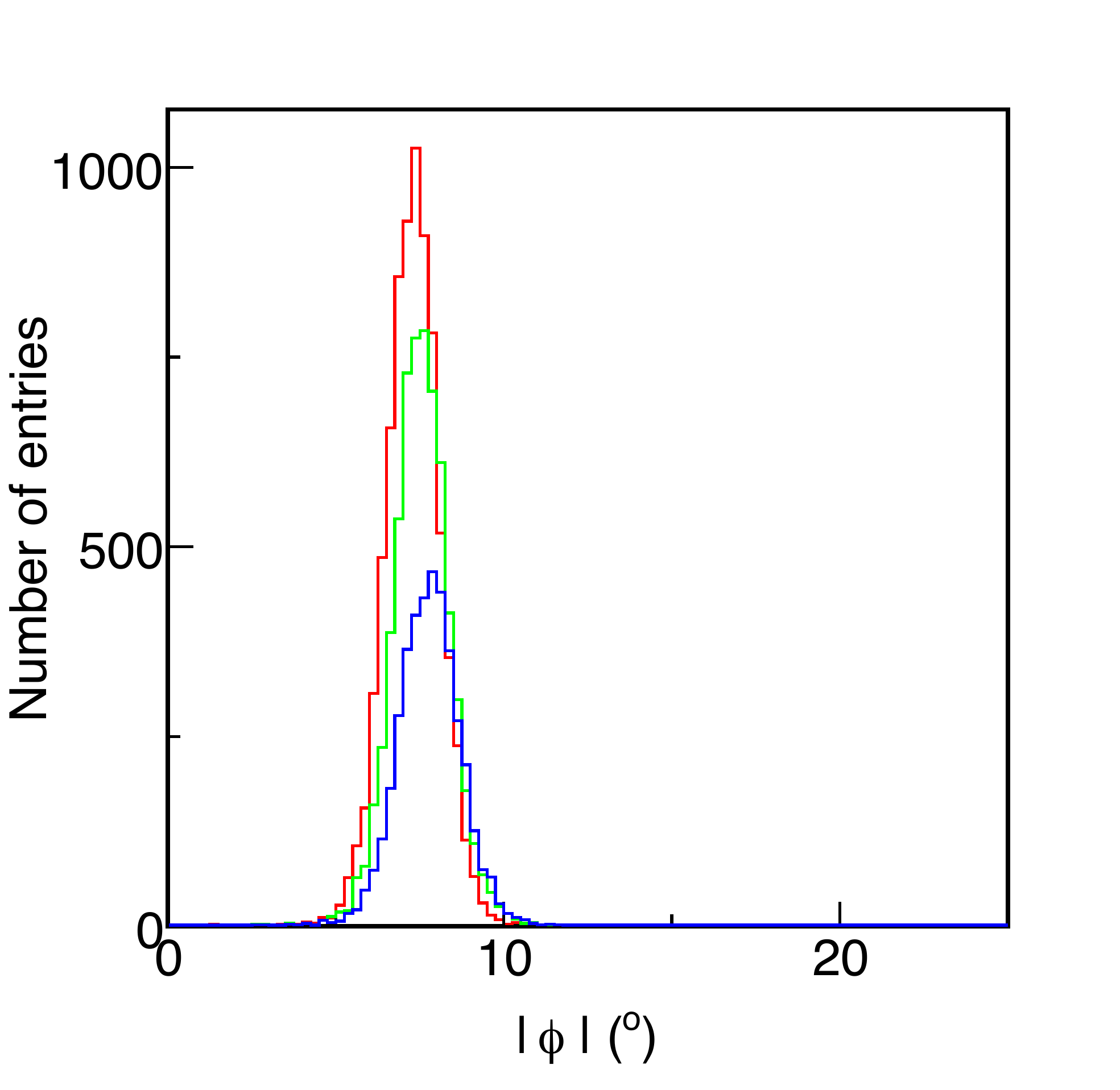}
\caption{Histograms of the absolute value of $\phi$ for alpha tracks from the shaded boxes in Fig.~\ref{fig:source_phi_vs_impact_parameter}. All peaks have a Gaussian width around 0.9 degrees, which is the worst-case limit for the angle measurement precision.}
\label{fig:source_absolute_phi}
\end{figure}

Fig.~\ref{fig:source_absolute_phi} shows the $\phi$ distributions for the standardized alpha tracks from the three sources. The spread of each distribution is due to a combination of non-zero source width and non-zero $y_{\textrm{cross}}$ window width in addition to the angular resolution we are trying to measure. We find that the Gaussian width of these peaks is 0.9 degrees for all three sources, indicating that our angle measurement precision is probably significantly better than 1 degree. The alpha tracks used here have a mean length of approximately 2.1\,cm. We can obtain a result that can be compared against the performance objectives by scaling to 1\,cm alpha track segments using Equation~5 in Ref.~\cite{Vahsen:2014fba}. The result is an estimated angular resolution of approximately 2.5 degrees, which is far better than the 15 degree objective.

\subsubsection{Ionization energy resolution}
\begin{figure}
\centering
\includegraphics[width=\columnwidth]{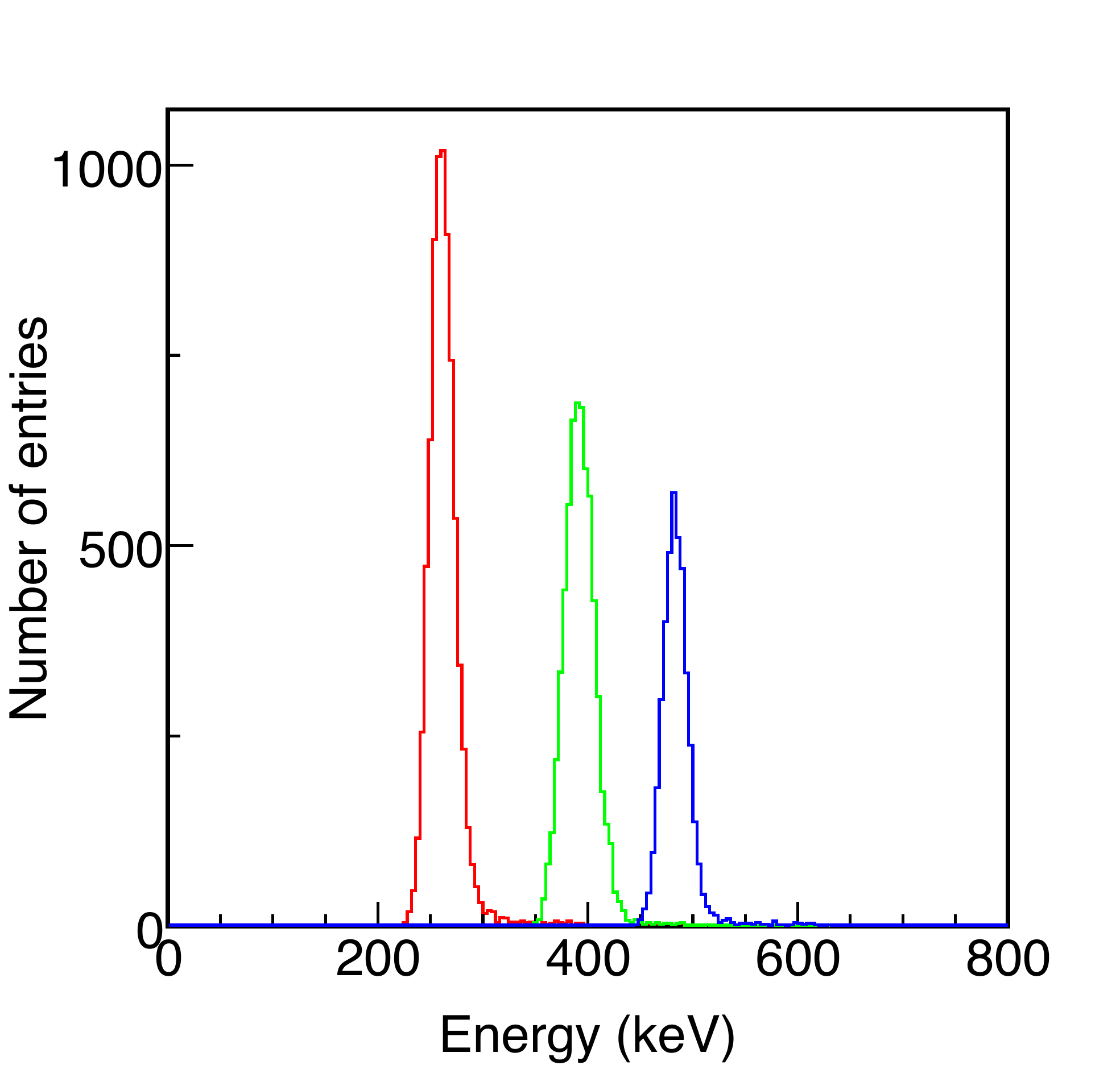}
\caption{A histogram of the total recovered ionization energy for alpha tracks from the top (left), middle (center), and bottom (right) calibration sources, chosen from the shaded boxes in Fig.~\ref{fig:source_phi_vs_impact_parameter}. With these selections, the difference between the recovered energy for the three sources is largely due to charge diffusion leading to unrecovered charge in pixels that do not cross threshold.}
\label{fig:source_energy_distributions}
\end{figure}

While the energy resolution measurements with the Fe-55 source (Section~\ref{subsubsection:gain}) characterize the detector performance for small (keV-scale) ionization signals at gains above 10,000, when operating the TPCs as neutron detectors we typically detect much larger signals and utilize much lower gains. The standardized alpha tracks allow us to estimate the energy resolution in this latter context. The total deposited ionization energy in one alpha track is the integral of the portion of the Bragg curve that subtends the sensitive volume. For the standardized alpha tracks, the tight angular selections also will select a very narrow band of deposited energies. Therefore we look at the total detected ionization energy for these tracks (Fig.~\ref{fig:source_energy_distributions}) and interpret the width of each peak as a worst-case limit on the energy resolution of the TPCs, for detected ionization energies of several hundred keV, at a gain of $1500$. The peak detected energy depends on the source $z$ due to the interaction of pixel threshold with diffusion. The RMS widths of the peaks are 6.4\% (top source), 4.3\% (middle source), and 3.7\% (bottom source) of the mean energy. This confirms that the energy resolution is excellent and improves with increasing energy.

\subsection{Testbeam}\label{sec:dd_generator}
In order to test the TPC and readout technologies in a realistic environment, we performed testbeam studies with both the prototype and production TPCs at a D-D fusion neutron generator~\cite{dd,dd2,doi:10.1063/1.4981896}. The generator produces a high rate (up to 10$^7$/s) of 2.5~MeV neutrons and copious X-rays, simulating both the target signal and most important physics background expected in beam background runs at SuperKEKB. In this section we focus on results from the prototype detector; we observed similar performance with the production TPC. The firmware trigger veto had not been developed by the testbeam and instead we used a hardware veto that was not 100$\%$ efficient, meaning that we accepted a large number of triggers from X-rays. 

\subsubsection{Testbeam description}\label{sec:dd_description}
\begin{figure}
\centering
\includegraphics[width=\columnwidth]{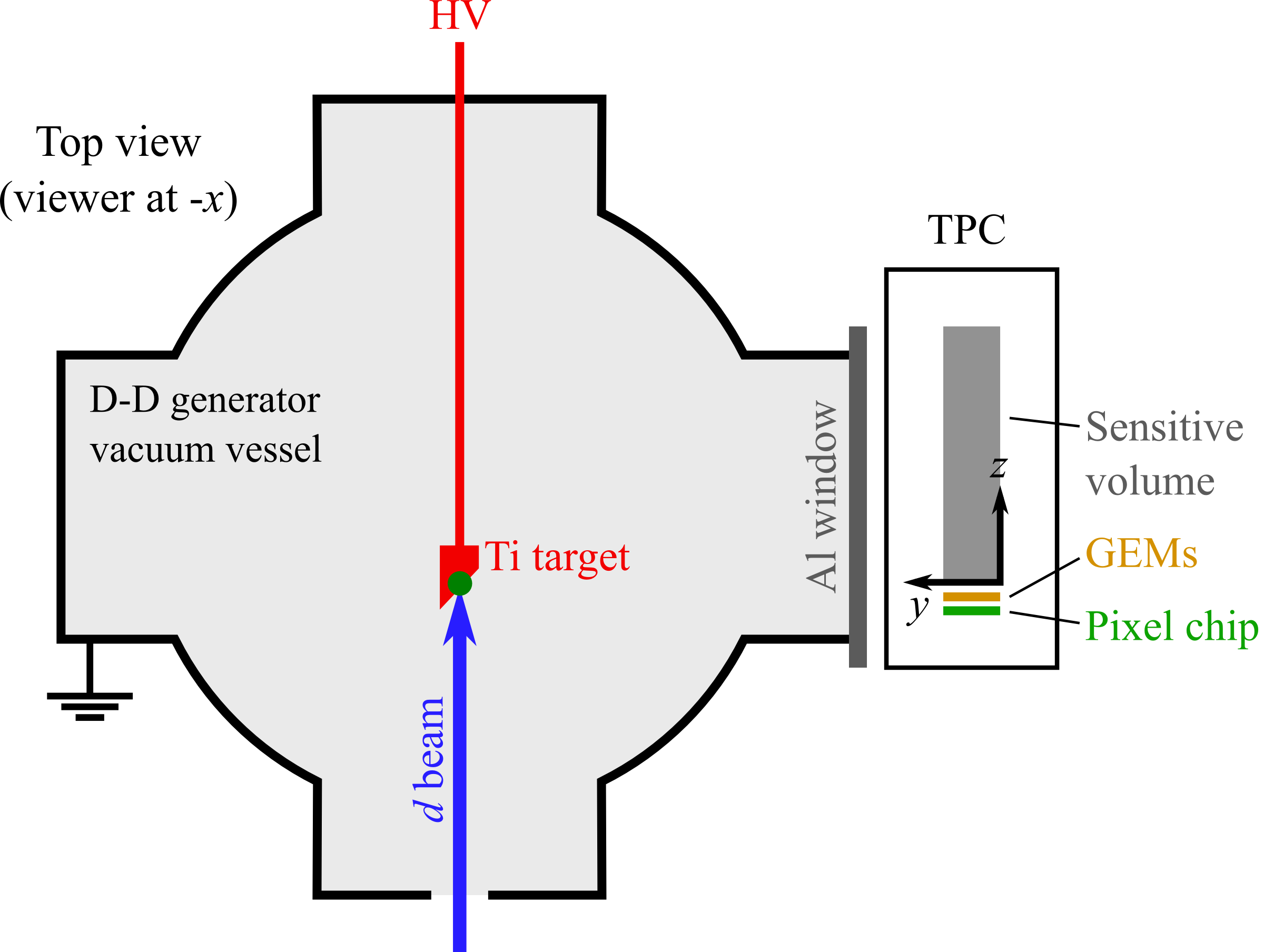}
\caption{A simplified rendering of the D-D testbeam layout as seen from above. The D-D generator vacuum vessel is on the left and the TPC vessel is on the right, with the $z$ axis of the TPC parallel to the deuteron beam.}
\label{fig:dd_setup}
\end{figure}

The D-D testbeam setup is illustrated in Fig.~\ref{fig:dd_setup}. A beam of deuterons accelerates through a variable accelerating potential (up to 90~kV) onto a fixed Ti target. The accelerated deuterons collide with stationary deuterons captured on the surface of the target and fuse, releasing neutrons in approximately half of the fusion events. We position the TPC so that the drift field is parallel to the deuteron beam. The sensitive volume of the TPC subtends polar angles ($\theta$) of 67-90 degrees with respect to the target and beam axis and 3.2 degrees in azimuth ($\phi$). 

\subsubsection{Testbeam results: rates}\label{sec:dd_rates}
\begin{figure}
\centering
\includegraphics[width=\columnwidth]{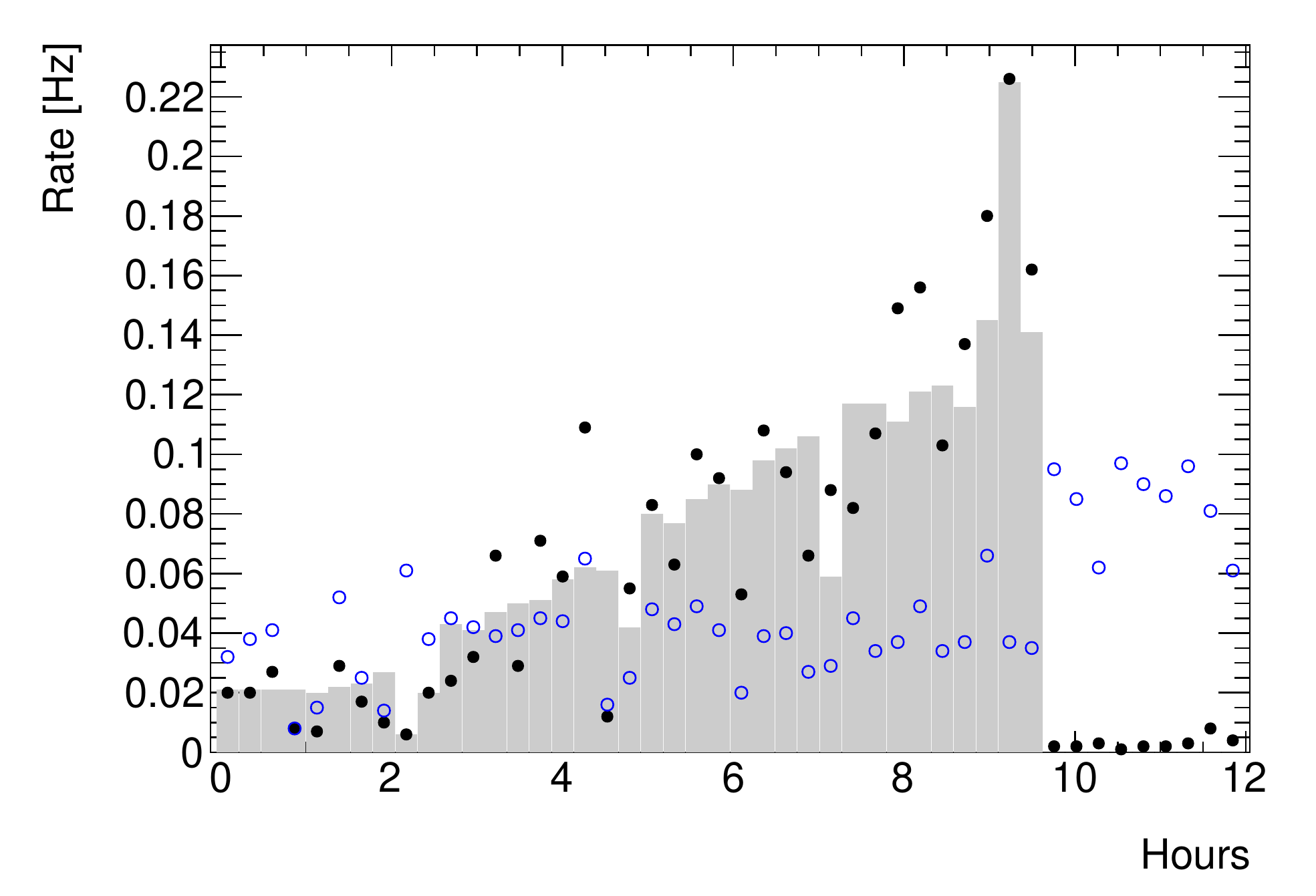}
\caption{(color online) Rates of neutron candidates (black closed circles) and good alpha candidates (open blue circles) compared to the scaled D-D neutron rate (gray) in a prototype TPC during D-D testbeam.}
\label{fig:neutron_rate}
\end{figure}

The rate of fast neutron production by the D-D generator is estimated using a $^3$He proportional counter embedded in a moderator (Health Physics Instruments, Model 6060). The neutron yield increases with the D$^+$ beam energy. In Fig.~\ref{fig:neutron_rate} we show the detected rate in the TPC and the rate measured by the $^3$He counter scaled to match the TPC rate as well as the rate of identified calibration alpha tracks during a 10-hour slow ramp-up of the accelerating potential and subsequent 2-hour pedestal period. We observe a good agreement between the shape of the generated neutron yield and detected neutron candidate curves. 

We also observe that the alpha rate is suppressed by a factor of roughly 50\% while the generator is on, likely caused by high trigger occupancy due to X-ray events. The neutron production rate is more sensitive to accelerating potential (roughly a factor of 10 increase per doubling) than the bremsstrahlung X-ray production (roughly a factor of 4 increase per doubling), perhaps explaining why the suppression factor of the alphas does not appear to depend on the neutron rate. 

\subsubsection{Testbeam results: uniformity}\label{sec:dd_uniformity}
\begin{figure}
 \begin{center}
  \includegraphics[width=1\columnwidth]{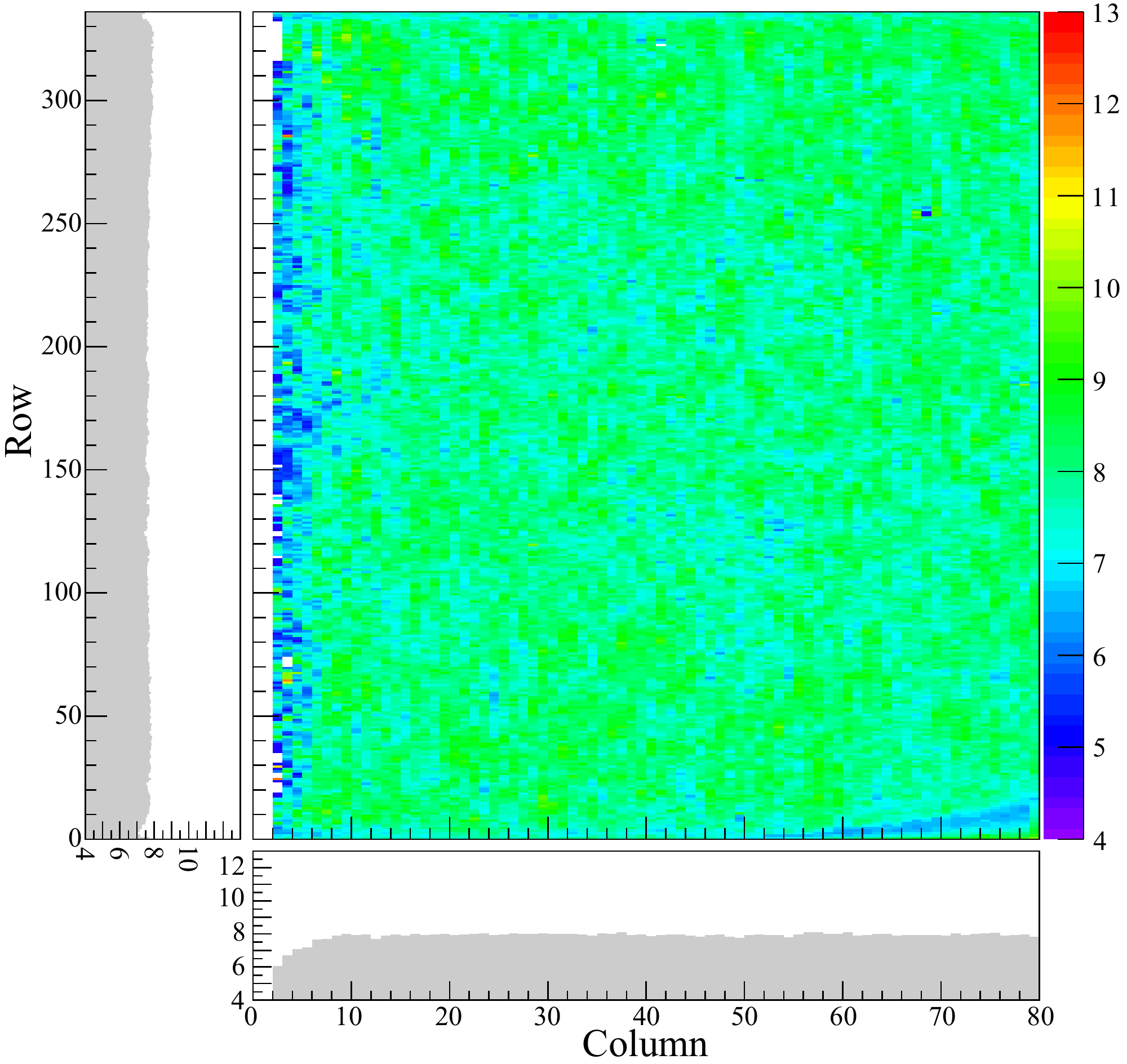}
  \caption{Charge scale uniformity illustrated using a large sample of X-ray and neutron recoil events in a prototype TPC using D-D testbeam data. The bin value is the average TOT of all hits from this sample plotted by row and column on the chip. The suppressed values on the left side of the chip are selection effects due to an alpha veto.}
  \label{fig:dd_uniformity}
 \end{center}
\end{figure}

The D-D testbeam dataset includes a large number of events uniformly distributed throughout the sensitive volume of the TPC, comprised largely of electron recoils from X-rays and nuclear recoils from neutrons. We use this sample to probe the combination of charge drift, amplification, transfer, detection, and calibration across the surface of the chip, which we call the charge scale uniformity. To select appropriate tracks to include in this sample, we first reject alpha tracks from the calibration sources by ignoring events that cross the side of the chip nearest the source and one or two other sides. We also reject events of non-physics origin (such as noise) by requiring the number of pixels above threshold in the event to be at least 5. We include all remaining tracks. For each pixel, we sum the TOT of every unsaturated hit in that pixel from all tracks in the sample, ignoring hits with TOT$<$5. We divide this by the number of hits in the same pixel in the entire sample, to obtain the average charge per hit, with the charge measured in units of TOT. We show the resulting charge scale uniformity in Fig.~\ref{fig:dd_uniformity}. To calculate the effective uniformity, we ignore the five columns closest to the calibration source and the outer two rows/columns on the remaining sides of the chip. We find that the mean TOT in this region is 8.0 with a standard deviation of 0.50, for a uniformity of 94\%.

\subsubsection{Testbeam conclusions}\label{sec:dd_conclusions}
We have concluded testbeam studies with one prototype and one production TPC at a D-D fast neutron generator. In both cases we observed stable operation in an environment comparable to SuperKEKB during operation with modest suppression of event rates due to trigger saturation (roughly 50\%) and excellent uniformity (94\%). With a fully efficient firmware trigger veto we expect little or no event rate suppression due to trigger saturation. These tests validate our TPC and data acquisition technologies for beam background measurements at SuperKEKB.

\subsection{Magnetic field test}
In June 2017, before installation of the final TPCs into the Belle II detector, we performed a final safety check to verify that the TPCs can operate in a strong magnetic field. We installed one final TPC in a large-bore electromagnet in the KEK North Counter Hall. The magnetic field was slowly increased from zero to 1.5\,Tesla, while we monitored the signal from two alpha calibration sources. The drift field and magnetic field directions were at right angles, so that we expected an observable shift in the alpha particle distribution on the pixel chip due to the magnetic field. We observed this shift, with the shift magnitude increasing with the magnetic field strength as expected. No adverse effects from the magnetic field were observed, and the detector remained fully functional and efficient both during and after this test.

\section{Discussion of results and wider impacts}

We have reported on the design, production, testing, and performance of the BEAST TPCs, a new generation of 3D directional nuclear recoil detectors. These detectors represent a factor of 60 (prototype) to 40 (final detector) scale-up in target volume over previous-generation detectors \cite{Vahsen:2014fba}, achieved by introducing a field cage, switching to the larger-area ATLAS FE-I4B pixel readout ASIC, and by developing custom DAQ firmware. We observe excellent uniformity in charge collection, resolving problems in previous-generation detectors. The low-level detector performance has been characterized with alpha particle calibration sources. We observe excellent angular resolution, energy resolution, and long term stability, and the detectors satisfy all design goals (Table\,\ref{tab:criteria}). The detectors have been demonstrated to operate stably in high magnetic fields and high neutrons fluxes. 

The BEAST TPCs were specifically optimized for the directional detection of (higher-energy) fast neutrons, and were operated at low gain in order to maximize operational stability, rather than sensitivity, in the high-background environment at SuperKEKB. Nevertheless, the detector sensitivity greatly exceeded our expectations. The high signal-to-noise and level of detail in measured 3D ionization distributions (Fig.~\ref{fig:event_displays}) will allow precision studies of nuclear recoils, and allow us to improve their simulation, in the future. Improved simulation of the straggling, length, and ionization distribution of low-energy nuclear recoils is critical for evaluating the sensitivity of other nuclear recoil detectors. For example, the ionization fluctuations of low-energy argon recoils are the largest uncertainty in the low-mass dark matter sensitivity of the DarkSide-50 experiment \cite{Agnes:2018ves}. Similarly, there has been a recent claim that keV-scale fluorine recoils are significantly longer than predicted by standard simulation tools~\cite{Tao:2019wfh}, which if correct, would drastically improve the sensitivity of gas-based directional dark matter detectors. Our previous generation detectors (which we operated at higher gain than the BEAST TPCs) \cite{Kim:2008zzi} appeared to have high single-electron sensitivity. BEAST TPCs operating at high gain could be used to characterizing the complete ionization distributions from low-energy nuclear recoils and would help clarify such issues.

To fully exploit the rich data and achieve optimal performance from the BEAST TPCs requires improved recoil-fitting algorithms, which we are currently developing. For this reason, the performance of recoil reconstruction and particle identification versus recoil energy will be published separately in the future. However, a few early results already exist, and offer a glimpse of what is possible: two recent PhD theses~\cite{phdthorpe,phdhedges} show that even at low gain, the detectors have a recoil energy threshold of order 10~keV. The TPCs also exhibit excellent particle identification capabilities via d$E/$d$x$ measurement for recoil energies above approximately 10-20~keV. This enables the identification of the species of the recoiling nucleus and the rejection of significant electron backgrounds. High efficiency and good particle identification capabilities at this energy scale are of interest for the detection of nuclear recoils in the context of dark matter searches. The high resolution charge readout in the BEAST TPCs, again at low gain, has also already been shown to enable full 3D fiducialization by measuring the transverse charge profile of tracks~\cite{Lewis:2014poa}. Such fiducialization is another desirable ingredient for improved background rejection in dark matter searches. As a result of these capabilities, gas TPCs with high-resolution charge readout have received strong interest from the directional dark matter detection community \cite{Battat:2016pap}. There has been a resurgence of interest in a large directional nuclear recoil detector, as such a detector could detect and distinguish coherent elastic neutrino-nucleus scattering from WIMP-nucleon scattering. For instance, recoils from solar neutrinos point (on average) back to the sun, while WIMP recoils are expected to point back to the constellation Cygnus. A directional detector could distinguish the two signals, and would have improved WIMP sensitivity below the so-called neutrino floor, while simultaneously allowing neutrino measurements \cite{CYGNUS}. An important next step towards demonstrating the feasibility of such a detector will be to explore the improved low-energy performance that can be achieved with the BEAST TPCs in high-gain (single-electron efficient) mode, and with a gas mixture fully optimized for dark matter searches. 

One common concern about scaling up highly segmented detectors to large sensitive volumes is the resulting high channel count and data rate, which can drive up cost and complexity. We note that the scheme demonstrated here --- thresholds applied at the pixel level before digitization, and a zero-suppressed, self-triggered readout --- results in a remarkably low data rate. The BEAST TPCs essentially only produce data when ionization is created in the sensitive volume. This is beneficial in two quite different contexts: First, in future low-rate applications with large sensitive volumes and readout planes with many pixel ASICs, combining region-of-interest self-triggers with a highly-multiplexed DAQ systems should allow high channel counts yet modest DAQ systems, reducing the cost of scaling up. Second, the self-trigger via firmware is beneficial in high-background applications. Vetoing electron-recoil events in the readout firmware, and not issuing a trigger for such events, was the key ingredient that allowed the BEAST TPCs to operate successfully in the high-radiation environment at SuperKEKB, without suffering from significant deadtime~\cite{Lewis:2018ayu}.

Finally, there is a broad need for improved fast-neutron detectors. The BEAST TPCs are compact and radiation hard, and simultaneously measure the location, 3D direction, and energy of neutron recoils with high accuracy. These capabilities should prove beneficial in a large number of contexts. Possible future applications of similar detectors include general characterization of neutron fields~\cite{Maire:2013era}, measurements in high-rate environments~\cite{Osipenko:2017gjd,Jocher:2019crv}, fast neutron imaging, nuclear security applications~\cite{Braverman:2018mcc}, and radiation monitoring in space~\cite{2007ess..book..765P,STAPELS2011342,Prettyman2011}.

\section{Acknowledgements}

We thank Kamaluoawaiku Beamer, Tommy Lam, Joshua Murillo, Blake Pollard, and Marc Rosen for their outstanding assistance during the design and production phases of the project. We are also grateful to Hiroyuki Nakayama for being our KEK liason, to Kodai Matsuoka for help with magnet operation, and to Karsten Gadow and Zachary Liptak for their help with TPC installation at KEK. Finally, We acknowledge support from the U.S. Department of Energy (DOE) via Award Numbers DE-SC0007852, DE-SC0010504, DE-AC02-05CH11231, and via the DOE-funded U.S. Belle II Project, administered by Pacific Northwestern National Laboratory.
\section*{References}
\bibliographystyle{elsarticle-num}
\bibliography{references}
\end{document}